\documentclass{emulateapj}

\usepackage{pslatex}
\newcommand{\vp}{v^\prime}
\newcommand{\Vp}{$v^\prime$}
\newcommand{\vpp}{$v\arcsec$}
\newcommand{\Jpp}{$J\arcsec$}
\newcommand{\Jp}{$J\arcmin$}

\newcommand{\Htwo}{\mbox{H$_{2}$}}
\newcommand{\Htwop}{\mbox{H$_{2}^{+}$}}

\newcommand{\xsig}{$X\,^1\Sigma_{g}^{+}$}
\newcommand{\xsigp}{$X\,^2\Sigma_{g}^{+}$}
\newcommand{\bsig}{$B\,^1\Sigma_{u} ^{+}$}
\newcommand{\bsigp}{$B^{\prime}\,^1\Sigma_{u} ^{+}$}
\newcommand{\bsigpp}{$B^{\prime\prime}\,^1\Sigma_{u} ^{+}$}
\newcommand{\cpi}{$C\,^1\Pi_{u} $}
\newcommand{\dpi}{$D\,^1\Pi_{u} $}
\newcommand{\dpip}{$D^{\prime}\,^1\Pi_{u} $}
\newcommand{\dpipp}{$D^{\prime\prime}\,^1\Pi_{u} $}
\newcommand{\kms}{km~s$^{-1}$}

\newcommand\bright{ergs cm$^{-2}$ s$^{-1}$ \AA$^{-1}$ arcsec$^{-2}$}
\newcommand\brightnoaast{ergs cm$^{-2}$ s$^{-1}$  sr$^{-1}$}
\newcommand{\ergs}{ergs cm$^{-2}$ s$^{-1}$ \AA$^{-1}$}
\newcommand{\lam}{$\lambda$}
\newcommand{\dlam}{$\lambda\lambda$}
\newcommand{\HST}{{\em HST}}
\newcommand{\IUE}{{\em IUE}}
\newcommand{\FUSE}{{\em FUSE}}
\newcommand{\lya}{{Ly$\alpha$}}
\newcommand{\lyb}{{Ly$\beta$}}
\newcommand{\ha}{{H$\alpha$}}
\newcommand{\hb}{{H$\beta$}}
\newcommand{\hg}{{H$\gamma$}}
\newcommand{\hd}{{H$\delta$}}
\newcommand{\lyc}{Lyc}

\journalinfo{Accepted for publication in the Astrophysical Journal}
\submitted{Received 2006 November 10; accepted 2007 January 5 }

\shorttitle{Hot Molecular Hydrogen in M27}
\shortauthors{McCandliss  et al.}

\begin{document}

\title{Molecular and Atomic Excitation Stratification in the Outflow of the Planetary Nebula M27}


\author{Stephan R. McCandliss\altaffilmark{1}, Kevin France\altaffilmark{2}, Roxana E. Lupu\altaffilmark{1}, Eric B. Burgh\altaffilmark{3}, Kenneth Sembach\altaffilmark{4}, Jeffrey Kruk\altaffilmark{1}, B.-G. Andersson\altaffilmark{1}, and Paul D. Feldman\altaffilmark{1}  }
\email{stephan@pha.jhu.edu}





\altaffiltext{1}{Department of Physics and Astronomy,
The Johns Hopkins University,
Baltimore, MD  21218.}
\altaffiltext{2}{Canadian Institute for Theoretical Astrophysics,  University of Toronto, Toronto ON M5S 3H8}
\altaffiltext{3}{Space Astronomy Laboratory,  University of Wisconsin - Madison, 1150 University Avenue, Madison, WI 53706}
\altaffiltext{4}{Space Telescope Science Institute, Baltimore, MD 21218.}


\begin{abstract}

High resolution spectroscopic observations with \FUSE\ and \HST\ STIS
of atomic and molecular velocity stratification in the nebular outflow
of M27 challenge models for the abundance kinematics in planetary
nebulae.  The simple picture of a very high speed ($\sim$ 1000 \kms),
high ionization, radiation driven stellar wind surrounded by a slower
($\sim$ 10 \kms) mostly molecular outflow, with low ionization and
neutral atomic species residing at the wind interaction interface, is
not supported by the M27 data.  We find no evidence for a high speed
radiation driven wind.  Instead there is a fast (33 -- 65 \kms) low
ionization zone, surrounding a slower ($\lesssim$ 33 \kms) high
ionization zone and, at the transition velocity (33 \kms),
vibrationally excited \Htwo\ is intermixed with a predominately neutral
atomic medium.   The ground state \Htwo\ ro-vibrational population
shows detectable absorption from \Jpp\ $\lesssim$ 15 and
\vpp\ $\lesssim$ 3.  Far-UV continuum fluorescence of \Htwo\ is not
detected, but Lyman $\alpha$ (\lya) fluorescence is present.  We also
find the diffuse nebular medium to be inhospitable to molecules and
dust.  Maintaining the modest equilibrium abundance of
\Htwo\ ($\frac{N(H_2)}{N(HI)} \ll $ 1) in the diffuse nebular medium
requires a source of \Htwo, mostly likely the clumpy nebular medium.
The stellar spectral energy distribution shows no signs of reddening
($E(\bv) < $ 0.01), but paradoxically measurements of
\ha/\hb\ reddening found in the literature, and verified here using the
{\it APO} DIS, indicate  $E(\bv) \sim $ 0.1.  We argue the apparent
enhancement of \ha/\hb\ in the absence of dust may result from  a two
step process of \Htwo\ ionization by  Lyman continuum (\lyc) photons
followed by dissociative recombination (\Htwo\ + $\gamma \rightarrow $
\Htwop\ + $e^- \rightarrow H(1s) + H (nl)$), which ultimately produces
fluorescence of \ha\ and \lya.  In the optically thin limit at the
inferred radius of the velocity transition we find dissociation of
\Htwo\ by stellar \lyc\ photons is an order of magnitude more efficient
than spontaneous dissociation by far-UV photons.  We suggest that the
importance of this \Htwo\ destruction process in \ion{H}{2} regions has
been overlooked.

\end{abstract}





\section{Introduction}

\begin{figure*}
\centerline{\includegraphics[]{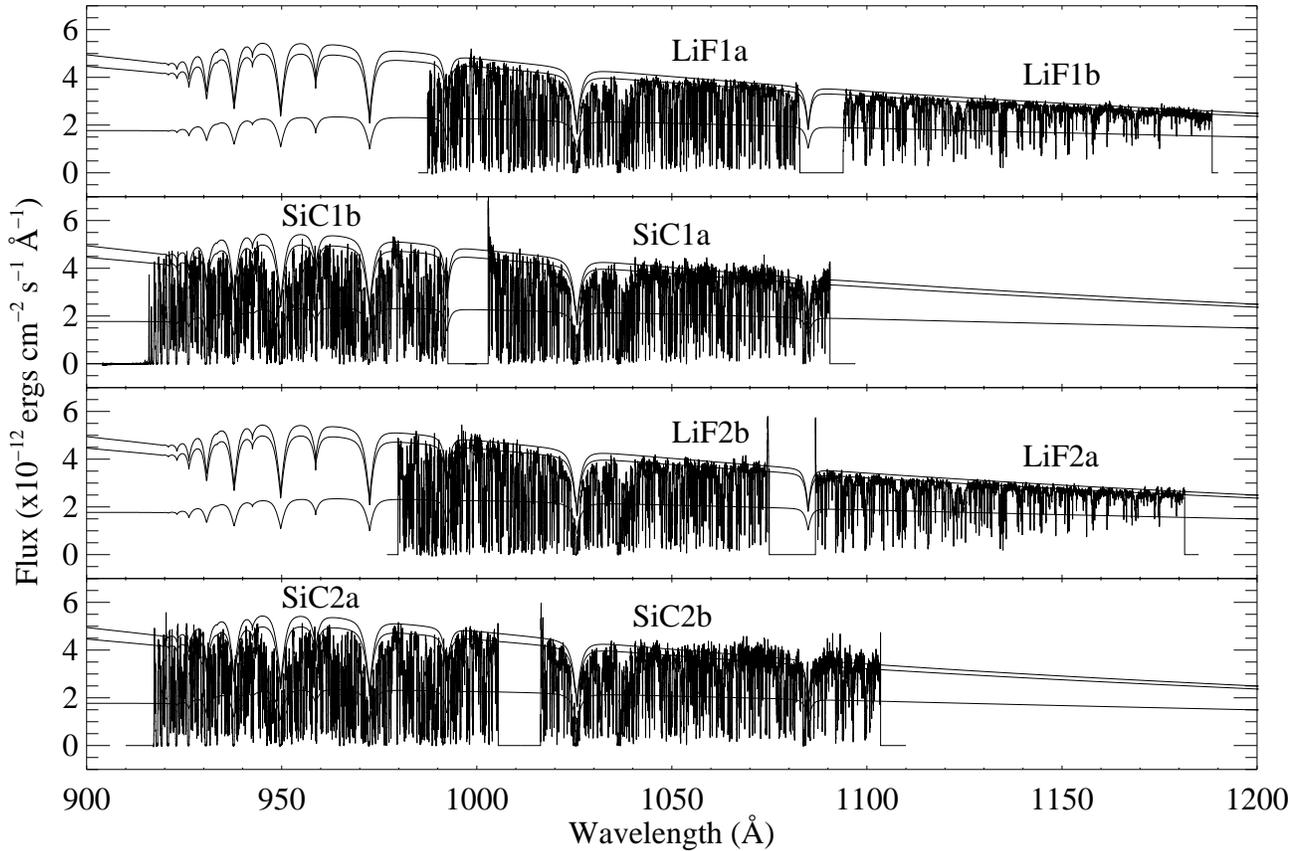} \\[12mm] }
\figcaption[f1.eps]{\label{edge} 
Flux of the orbital coadds for the eight spectral
segments taken through the HIRS spectrograph aperture.  Individual segments are labeled. Overplotted is the model SED without extinction (upper line) and with  $E(\bv)$ = 0.005 mag (slightly lower line) and $E(\bv)$ = 0.05 (lowest line) for a standard galactic extinction curve.  $E(\bv)$ = 0.005 mag provides the best match to the SED in the SiC channels.  }
\end{figure*}

M27 (NGC~6853, the Dumbbell) exhibits a bi-polar morphology, often associated with molecular hydrogen (\Htwo) infrared
emission in planetary nebulae
(PNe) \citep{Kastner:1996}.  High resolution spectra of the
hot central star (CS), acquired with the Far Ultraviolet Spectroscopic
Explorer (\FUSE), have revealed an unusually rich set of narrow
\Htwo\ absorption features spanning the entire spectral
bandpass, an indication that the molecule is vibrationally excited.
\FUSE\ carries no on-board source for wavelength calibrations and
consequently M27 has been observed numerous times for this purpose,
resulting in a high signal-to-noise data set \citep[see Figure~\ref{edge} and][]{McCandliss:2007}.  The usefulness of
these lines for wavelength calibration aside, we are presented with an
interesting puzzle.  What physical processes excite \Htwo\
in PNe and how does it survive in this high temperature and highly
ionized environment?

At first glance it seems surprising to find \Htwo\ in PNe at
all. For a typical electron density of 300 cm$^{-3}$ and temperature of 10,000
K, we estimate an e-fold lifetime $\approx$ 230 years, assuming
electron impact dissociation of \Htwo\ at the rate given by
\citet[][$\gamma_{e^-}$ = 4.6 $\times$ 10$^{-13}$ cm$^{3}$
s$^{-1}$]{Martin:1998}.  Nevertheless there is a whole class of PNe
with bi-polar morphology in which infrared \Htwo\ emission is the
defining characteristic \citep{Kastner:1996,Zuckerman:1988}.
\citeauthor{Zuckerman:1988} concluded that the infrared H$_2$ emission in
M27 is shock excited, based on IR spectroscopic diagnostics.  This
result is considered somewhat surprising because the close presence of
a hot star suggests that far-UV continuum excited fluorescence of the
\Htwo\ may be important.  For instance, \citet{Natta:1998}
calculate that continuum pumped fluorescence processes dominate thermal processes in PNe evolutionary models at late times $\gtrsim$ 5,000 yrs.

\citet{Herald:2002, Herald:2004} and \citet{Dinerstein:2004} have
reported excited \Htwo\ absorption features in \FUSE\ spectra
of a handful of CSPN located in the Galaxy
and Large Magellanic Cloud.  \citet{Dinerstein:2004} and
\citet{Sterling:2005} have used \FUSE\ to examine PNe with
strong, extended \Htwo\ infrared (IR) emission.  They find
the detection of IR emission is no guarantee for finding
excited \Htwo\ in absorption and conclude the
molecular material in these systems is clumped.

Many authors have discussed the clumped structures and dense knots
observed in PNe, along with evidence for their origin and chemical
composition \citep[c.f.][and references therein]{O'Dell:2002,
O'Dell:2003, Huggins:2002,  Bachiller:2000, Cox:1998, Meaburn:1993,
Reay:1985}.  These structures are isolated photodissociation regions
(PDR) immersed in  \ion{H}{2} regions, and it is natural to expect them
to be reservoirs of molecular material in PNe, regardless of the
mechanism that causes them to form.  \citet{Capriotti:1973} studied
the dynamical evolution of a radiation bounded PN and found that dense
neutral globules form as the result of a gradually weaking radiation
field in an outward propagating ionization front.  A Rayleigh-Taylor
like instability develops and eventually an optically thick spike of
neutral material becomes separated from the front and forms a high
density globule where  molecules can presumably form.   The formation
of \Htwo\ at later stages,  on dust grains and in the high electron
density environment via the H + H$^-$ $\rightarrow$ H$_2$ + e$^-$
reaction, has been modeled by \citet{Aleman:2004} and
\citet{Natta:1998}.  \citet{Williams:1999} has discussed a shadowing
instability, somewhat  similar to that of \citet{Capriotti:1973}, where
small inhomogeneities perturb a passing supersonic ionization front,
causing a corrugation that produces a large neutral density contrast
further downstream.  \citet{Dyson:1989} suggested molecular knots in PN
are relic SiO maser spots, which originate in the atmosphere of the
asymptotic giant branch (AGB) progenitor.  Others have proposed that
molecules may also originate from relic planetary material, either
accreted and then ejected or swept up during the AGB phase preceding
the formation of the nebula \citep[c.f.][]{Wesson:2004, Rybicki:2001,
Siess:1999, Livio:1983}.  \citet{Soker:1999} has discussed the
signatures of surviving Uranus/Neptune-like  planets in PNe.

\citet{Redman:2003} have emphasized the importance of molecular
observations as a means to establish the evolutionary history of the
clumps, which should depend on when and where they formed during the
AGB $\rightarrow$ PN transition.  The expectation is for dense clumps
formed within an AGB atmosphere to have more complex molecules, due to
higher dust extinction and molecular shielding, than clumps formed
later in the PN phase when the stellar radiation field is harder and
the overall density lower.  The abundance of atomic and molecular
species with respect to \Htwo\ is of fundamental importance
to assessing the predictions for the chemical evolution in PNe clumps.
Observations of atomic and molecular excitation stratification in the
outflow provide a fossil record of the AGB $\rightarrow$ PN transition
containing clues to the origins of the clumps.

Towards this end we explore high resolution far-ultraviolet absorption
spectroscopy, provided by \FUSE\ and the Space Telescope Imaging
Spectrograph (STIS) on board the Hubble Space Telescope (\HST), to
reveal the nebular outflow kinematics imprinted on the line profiles of a
wide variety of molecular and atomic species, including H$_2$, CO,
\ion{H}{1}, \ion{C}{1} \ion{-}{4}, \ion{N}{1} \ion{-}{3}, \ion{O}{1},
\ion{O}{6}, \ion{Si}{2} \ion{-}{4}, \ion{P}{2} \ion{-}{5}, \ion{S}{2}
\ion{-}{4}, \ion{Ar}{1} \ion{-}{2} and \ion{Fe}{2} \ion{-}{3}.
Absorption spectroscopy samples material directly along the
line-of-sight.  We will sometimes refer to it as the diffuse nebular
medium to distinguish it from the extended clumpy medium,
offset from the direct line-of-sight, which undoubtedly differs in
physical and chemical composition.  

Our main objective is to quantify the \Htwo\
excitation state, its abundance with respect to \ion{H}{1}, and its
outflow velocity with respect to other emitting and absorbing atomic
and molecular species.  We use this information, along with the
spectral energy distribution (SED) of the central star, to constrain
\Htwo\ formation and destruction processes in the nebula.

The analysis is supported by a variety of ancillary data on M27.
Longslit far-ultraviolet spectroscopy from a JHU/NASA sounding rocket
(36.136 UG) provides an upper limit to the \Htwo\ continuum
fluorescence.  Longslit optical spectrophotometry with the Double
Imaging Spectrograph (DIS) at Apache Point Observatory (APO) yields the
absolute flux of the CS and the surrounding nebular Balmer line
emission.   Dwingeloo Survey 21 cm data helps constrain the velocity
structure of \ion{H}{1}.  We begin by reviewing previous observations
and physical properties of the nebula followed by a discussion of the
various data sets.  We then present the analysis, discussion,
conclusions and  suggestions for future work.

\section{Physical Properties and Previous Observations of M27}
\label{previous}

\begin{figure}
\centerline{\includegraphics[]{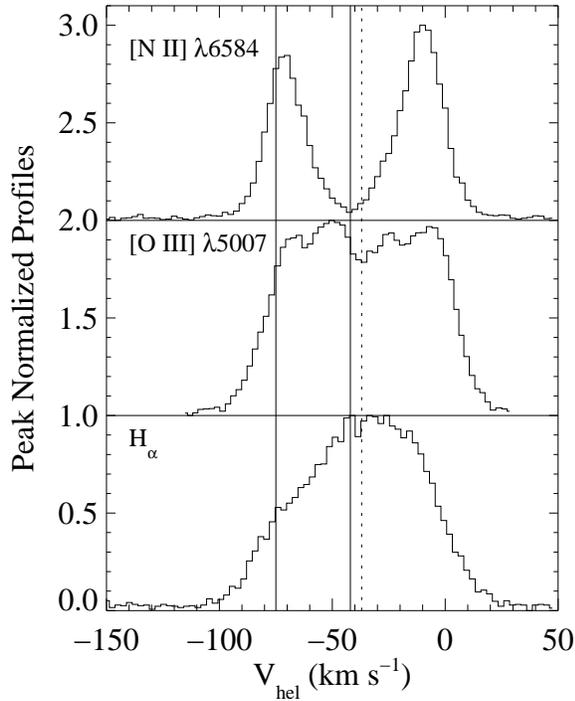}}
\figcaption[f2.eps]{ \label{meaburnvel}
Normalized intensity profiles of $H_{\alpha}$, [\ion{O}{3}] $\lambda$5007 and [\ion{N}{2}] $\lambda$ 6584 as a function of velocity.  $H_{\alpha}$ and
[\ion{N}{2}] $\lambda$ 6584 were digitized from \citet{Meaburn:2005} and
[\ion{O}{3}] $\lambda$5007 was digitized from \citet{Meaburn:1992}.} 
\end{figure}

M27 has an elliptical shape $\approx$ 8\arcmin $\times$
5\arcmin\ as seen in visual photographs \citep{Burnham:1978}.   Deep
narrow band CCD imaging of H$\alpha$ + [\ion{N}{2}] and [\ion{O}{3}] 
by \citet{Papamastorakis:1993} has shown a faint halo of these emissions
extending to 17\arcmin.  Its bi-polar morphology is manifest in H$\alpha$ images
as the ``Dumbbell''  and in the 2.12 $\mu$m image of
\citet{Kastner:1996} as a clumpy ``bow-tie'' aligned with the semi-minor axis. 
The ends of the bow-tie are $\approx$ 3\arcmin\ wide, the waist 
$\approx$ 1\arcmin\ and the length  $\approx$ 6\arcmin\ \citep[c.f.][]{Kastner:1996,
Zuckerman:1988}.

\citet{Napiwotzki:1999} determined the DAO CS parameters from NLTE
model atmosphere analysis and evolutionary considerations ($T$ =
108,600 $\pm$ 6800 K, $\log(g)$ = 6.7 $\pm$ 0.23,  $\log$(He/H)  by
number of --1.12 dex, and  $M_{*}$ = 0.56 $\pm$ 0.01 M$_{\sun}$).
Astrometric observations of the CS by \citet{Benedict:2003} produced a
distance of $d$ = 417$^{+49}_{-65}$ pc.  At this distance the $\approx$
8\arcmin $\times$ 5\arcmin\  ellipse is 1 pc $\times$ 0.6 pc and the
1025\arcsec\ diameter of the halo is 2.1 pc. \citeauthor{Benedict:2003}
give other physical parameters as well,  $V = $13.98$\pm$0.03, a total
extinction $A_V$ = 0.3$\pm$0.06 and a stellar radius $R_{*}$ =
0.055$\pm$0.02 R$_{\sun}$.  They find a bolometric magnitude
$M^{*}_{bol}$ = --1.67$\pm$0.37, which yields a luminosity for the
central star of $L_* = 366~L_{\odot}$. The radial velocity of the
central star in the heliocentric system was determined by
\citet{Wilson:1953} to be V$_{sys}$ = --42 $\pm$ 6~\kms.

\citet{Barker:1984} has reported on the analysis of line emission from
the nebula as derived from both ground based and International
Ultraviolet Explorer (\IUE) spectra.  The emission is typical of an
\ion{H}{2} region with an electron temperature of $\approx$ 10,000
K, an electron density of $\approx$ 300 $\pm$ 100 cm$^{-3}$,  and an
elevated metallicity with respect to solar in CNO.  These parameters
agree well with those found  by \citet{Hawley:1978} in a similar ground
based study of the nebula. \citet{Barker:1984} adopts a line reddening
parameter of $c = 0.17$ ($c$ is defined in \S~\ref{blr}).  The
literature reveals a range of nebular line reddening parameters,
0.03$\le c \le$ 0.18 \citep[c.f.][]{Miller:1973, Kaler:1976, Cahn:1976,
Barker:1984, Ciardullo:1999}.  This translates to 0.02$\le E(\bv) \le$
0.12, \citep[using  $c$/$E$(\bv) = 1.5,][Figure 4]{Ciardullo:1999}.
\citet{Pottasch:1977} derived $E(\bv)$ = 0.10 $\pm$ 0.04 by removing a
small (some would say imperceptible) 2175 \AA\ bump from an
\IUE\ spectrum.  This value is nearly the same as that given by
\citet{Harris:1997} and \citet{Benedict:2003}, based on trigonometric
distance determinations and absolute visual magnitude considerations.
\citet{Cahn:1992} report $c_r$ = 0.04 derived from the ratio of 5 GHz
radio flux density to \hb\ absolute flux.

Velocity and position resolved emission line spectroscopy  by
\citet{Goudis:1978} showed faint [\ion{O}{1}] and [\ion{N}{2}] profiles
(beam size of 83\arcsec) taken near the CS to be double peaked
symmetrically around the radial velocity of the system.  In the outer
regions the profiles converged to single peaks at the systemic
velocity. The bright [\ion{O}{3}] lines (beam size of 30\arcsec) showed
similar behavior but with a lower double peaked splitting.  These
observations are consistent with the presence of two nested shells
expanding with projected velocities of $\approx$ 33 and 15
\kms\ respectively.  \citet{Meaburn:1992} used a fiber optic image
dissector fed into an echelle spectrograph to closely examine the
\ion{O}{3} velocity structure in the immediate vicinity of the CS
($\approx$ 36\arcsec\ North-South), and found quadruple peaked line
profiles consistent with gas expanding at velocities of 31 \kms\ and 12
\kms.  \citet{Meaburn:2005} showed that H$\alpha$ exhibits a filled-in
asymmetric profile with a half-width half-maxium of $\approx$~37 \kms,
in contrast to the double peaked emission exhibited by [\ion{N}{2}]
$\lambda$6584 in the vicinity of the CS.  \ion{He}{2} $\lambda$6560
shows a much narrower profile and is consistent with turbulent
broadening alone.  Examples of the [\ion{N}{2}] $\lambda$6584 and
H$\alpha$ profiles from \citet[][Figure 3]{Meaburn:2005} are reproduced
in the top and bottom panels of Figure~\ref{meaburnvel}.  The middle
panel of Figure~\ref{meaburnvel} shows \ion{O}{3} \lam5007 taken from
of \citet[][Figure 6d]{Meaburn:1992}.  Note  the base of these emission
features extend to $\approx \pm$ 60 \kms.

\begin{figure*}[t]
\centerline{\includegraphics[]{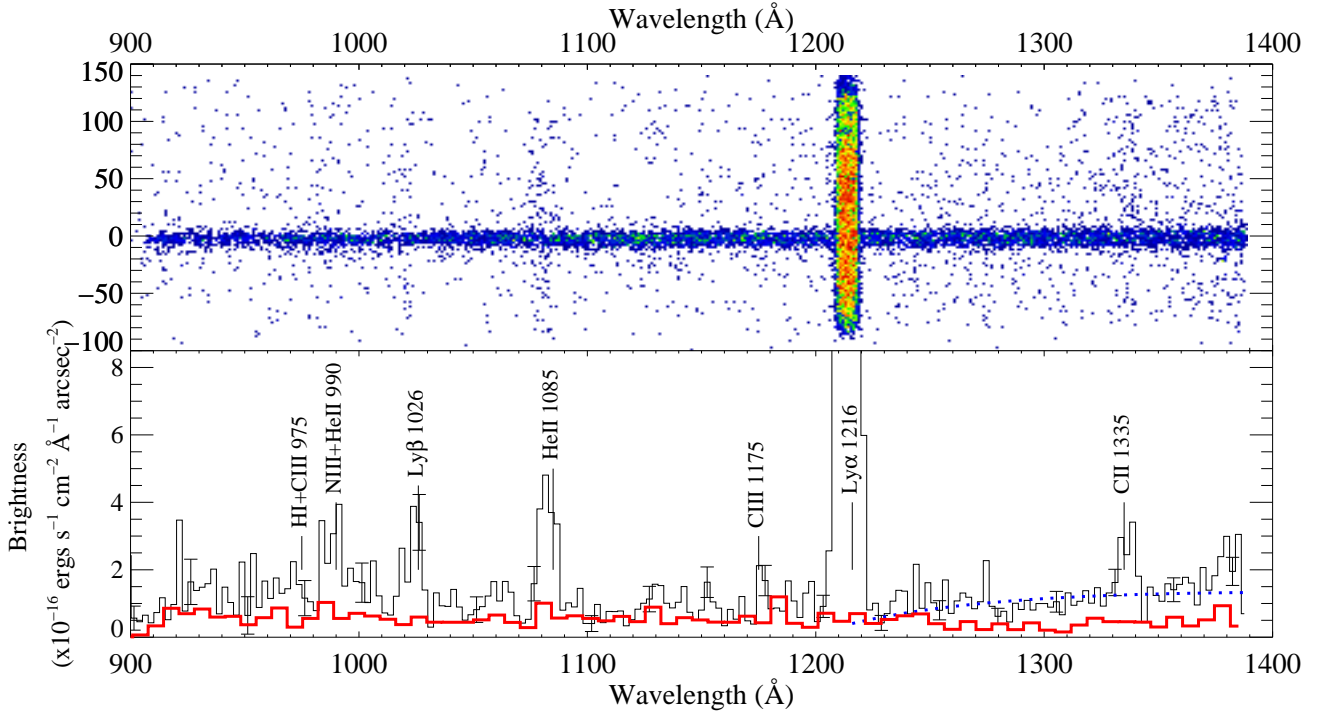}}
\figcaption[f3.eps]{ \label{figrock}
Top  -- raw longslit far-UV spectrum from sounding rocket experiment 36.136 UG.  Bottom -- total nebular brightness per \AA\ summed over the whole slit, excluding the continuum source.  The red line is the estimated detector background.  There
is no significant dust scattering or \Htwo\ continuum fluoresence emission above this background.  The rise longward of \lya\ is likely caused by
2s -- 1s,  2-photon emission of recombining hydrogen.  The dotted blue line shows the expected brightness for a column of $N(2s)$ = 4 $\times$ 10$^{8}$ cm$^{-2}$.
}
\end{figure*}

Observations by \citet{Huggins:1996} of the CO(2--1)  230 GHz emission
lines at a position 68$\arcsec$W and 63$\arcsec$S from the CS also
exhibit similar double peaked line profiles, with a total separation
$\approx$ 30 \kms.  At this position they estimate a CO column density
of 6.9 $\times$ 10$^{15}$ cm$^{-2}$ (assuming LTE with 5 $< T_{ex} <
$150 K). At a position 10\arcsec\ due West of the CS,
\citet{Bachiller:2000} show lines split by 53 $\pm$ 1 \kms (their
Figure 4), yielding an expansion velocity of $\approx$ 27 \kms\ with
symmetry about a heliocentric velocity of --43 \kms (at M27, $V_{hel}$ = $V_{lsr}$ -- 17.8 \kms), in good agreement
with the Wilson systemic velocity. They present a map of the region
(beam size 12\arcsec), showing the CO emission to be clumpy and
more-or-less coincident with the molecular hydrogen infrared emission
observed by \citet{Kastner:1996}.  \citet{Bachiller:2000} present a
picture of an ionized central region with an electron density of about
$n_e \sim$ 100 cm$^{-3}$ surrounded by a ring of molecular clumps with
densities $\sim$ 10$^4$ cm$^{-3}$ undergoing photodissociation.  They
suggest the clumps are similar to the ``cometary'' shaped features
observed in great detail in the Helix nebula
\citep[c.f.][]{O'Dell:2002, Huggins:2002, Meixner:2005}, although the
cometary morphology is not as prominent in the Dumbbell as it is in the
Helix.  \citet{Meaburn:1993} first noted  \citep[see also \HST\ images
by][]{O'Dell:2002} that some clumps appear as dark knots against the
very bright \ion{O}{3} emission that surrounds the central region.

\citet{McCandliss:2001a} gave a preliminary analysis of  molecular
hydrogen absorption in the \FUSE\ spectra, identifing two absorption
systems, one blueward of the CS radial velocity and one redward. The
blueward  component has lines originating from an excited electronic
ground state (\xsig), with rotational levels (\Jpp) up to 15 and
vibrational levels (\vpp) at least as high as 3. Many lines originating
from \vpp~=~1~and~2 are located longward of the ground state bandhead
($\vp$ ~--~\vpp ~=~0~--~0) at 1108.12~\AA, and as such are unambiguous
markers of ``hot'' \Htwo.  In contrast,  the redward component is much
``cooler,'' showing no unusual excitation
(J$^{\prime\prime}_{max}~\approx$~4, \vpp~=~0).  It is associated with
non-nebular foreground gas.   Here we will refine the original
analysis, which was based on LWRS data acquired under \FUSE\ PI Team
Program P104.

Lastly we note, \lya\ fluorescence of \Htwo\ has recently
been found in the PNe  M27 and NGC 3132  by \citet{Lupu:2006}.  The
 fluorescence is a direct consequence of excited molecular
hydrogen, as the mechanism requires a significant population in the
\vpp~= 2 levels in the presence of strong \lya\ emission.  Herald
reports (private communication) that \FUSE\ observations of the CS of
NGC 3132 also exhibit excited \Htwo\ absorption lines with
a roughly thermal ro-vibrational distribution of $\approx$ 1750 K.

\section{Datasets}

\subsection{\FUSE}
\label{fusespec}

A description of the data set and processing procedures for the
\FUSE\ spectra of the CS (MAST object ID GCRV12336) used in this study
can be found in \citet{McCandliss:2007}.  For a description of the
detectors, channel alignment issues, systems nomenclature and other
aspects of the \FUSE\ instrument, see \citet{Moos:2000} and
\citet{Sahnow:2000}.  Briefly, the CS was observed numerous times for
observatory wavelength calibration purposes.  High signal-to-noise
spectra, acquired in time tag mode through the high resolution slits
(HIRS) of the eight different channel segments have been assembled.
Data for each extracted channel segment consists of three 1-D arrays: wavelength (\AA), flux (\ergs), and an estimated
statistical error (in flux units).  Figure~\ref{edge} shows the eight individual channel segments and their spectral range for all the coadded HIRS spectra.
The high density of the \Htwo\ features is apparent.

\begin{figure*}[t]
\centerline{\includegraphics[]{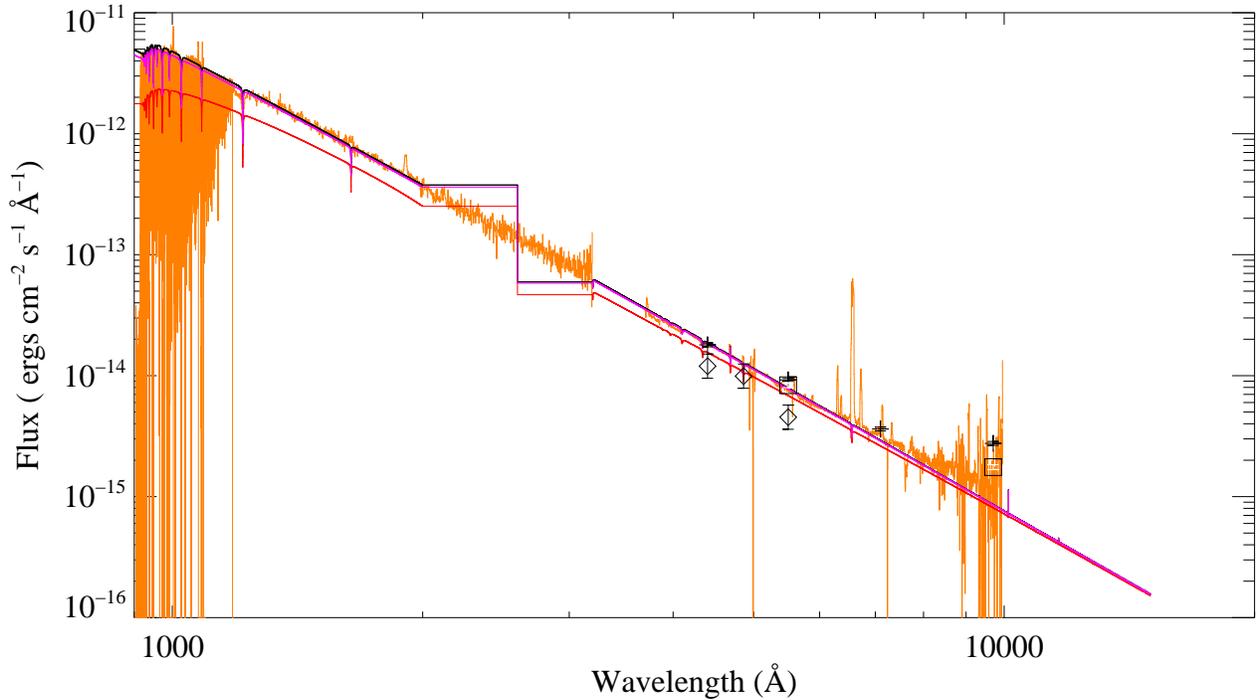}}
\figcaption[f4.eps]{ \label{sed}
Log-log plot of the M27 central star SED from 900 - 20000 \AA, as measured by \FUSE, \IUE, APO and the optical photometry of \citet{Tylenda:1991} ($\diamond$), \citep{Ciardullo:1999} (\sq) and  \citet{Benedict:2003} (+).  Overplotted in black, purple and red are the respective unextinguished, lightly extinguished ($E(\bv)$ = 0.005) and heavily extinguished ($E(\bv)$ = 0.05) stellar continuum models.  Break between 2000 and 3000 \AA\ indicates a gap in the stellar model. }
\end{figure*}

In the low sensitivity SiC channels
the continuum signal-to-noise is between  10 -- 25 per pixel, while in the
higher sensitivity LiF channels it is  20 -- 55.  These signal-to-noise
ratios are purely statistical and do not account for systemic errors,
such as detector fixed pattern noise.  The resolution of the spectra
changes slightly as a function of wavelength for each channel segment.
In modeling the \Htwo\ absorption (\S~\ref{cogsec}), we
find that a gaussian convolution kernel with a full width at half
maximum of 0.056 \AA\ at 1000 \AA\ (spectral resolution $R$ $\approx$ 18,000,
velocity resolution ${\delta}V$ $\approx$ 17 \kms) provides a good
match to the unresolved absorption features throughout most of the
\FUSE\ bandpass.

Close comparison of the wavelength registration of overlapping segments
reveals isolated regions, a few \AA\ in length, of slight spectral
mismatch ($\sim$ a fraction of a resolution element) in the wavelength
solutions.  Consequently, combining all the spectra into one master
spectrum will result in a loss of resolution.  However, treating each
channel/segment individually increases the bookkeeping associated with
the data analysis.  Further, because fixed pattern noise tends to
dominate when the signal-to-noise is high, there is little additional
information to be gained in analyzing a low signal-to-noise data set
when high signal-to-noise is available.  For these reasons we elected
to form two spectra, each of which covers the 900 -- 1190 bandpass
contiguously, using the following procedure.  

The flux and error arrays
for the LiF1a and LiF1b segments were interpolated onto a common linear
wavelength scale with a 0.013 \AA\ bin, covering 900 -- 1190 \AA.  The
empty wavelength regions, being most of the short wavelength region
from 1000 \AA\ down to the 900 \AA\ and the short gap region in between
LiF1a and LiF1b, were filled in with most of SiC2a and a small portion
of SiC2b respectively.  We refer to this spectrum as s12.  The LiF2b,
LiF2a, SiC1a and SiC1b segments were merged similarily into a spectrum,
s21.  Absorption line analyses were carried out using the composite s12 and s21
spectra (\S~\ref{cogsec}).\footnote{The s12 and s21 spectra are available through the
H$_2$ools website \citep{McCandliss:2003} along with the original complete M107
data product (LWRS, MDRS and HIRS) processed by Jeffrey Kruk (private
communication).}

\subsection{Far-UV Longslit Observations of M27 from JHU/NASA Sounding Rocket 
36.136~UG}

\label{rocket}

\begin{figure*}[t]
\centerline{\includegraphics[]{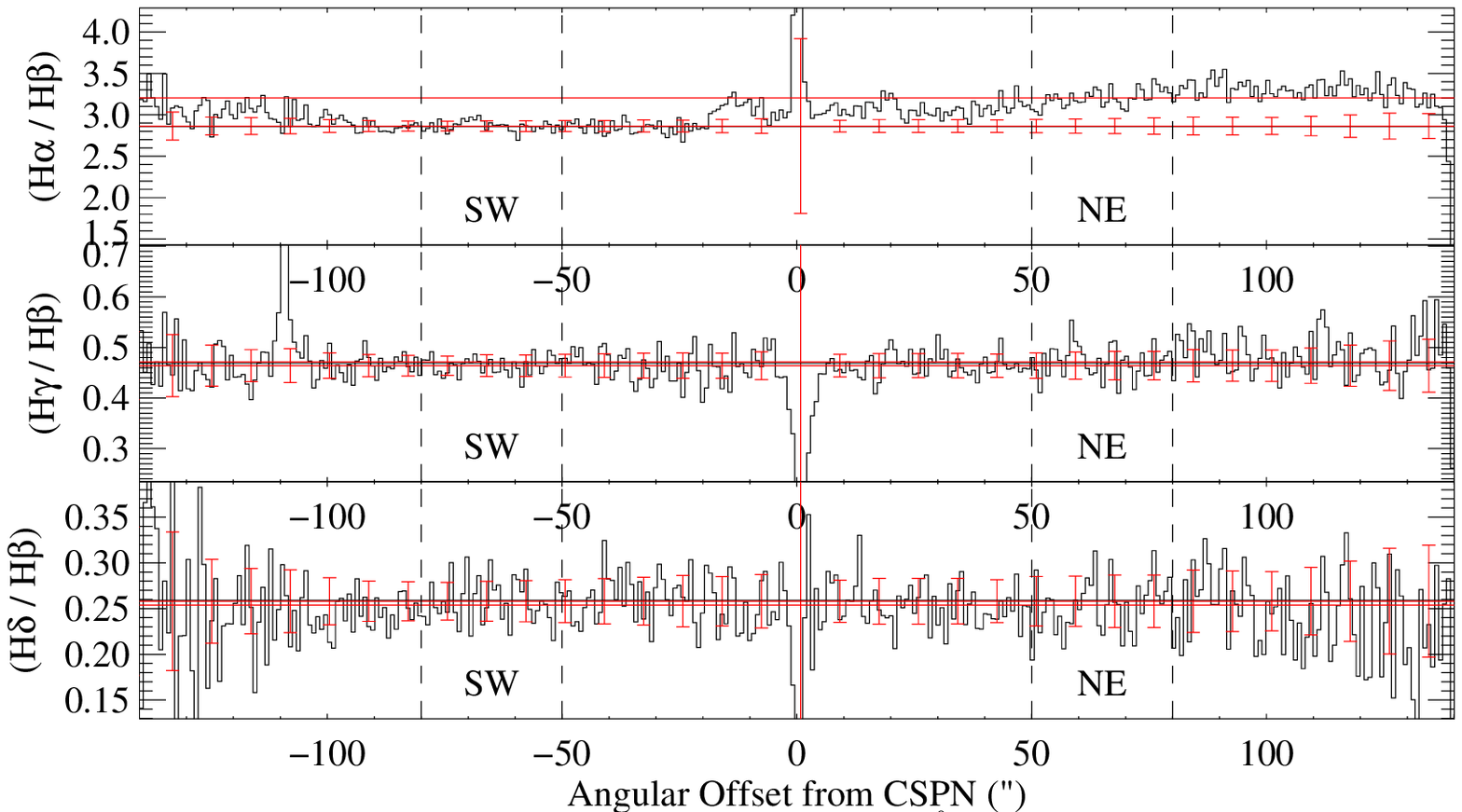} \\[10mm] }
\figcaption[f5.eps]{ \label{recomb}
Long slit profiles of \ha/\hb, \hb/\hg, and \hb/\hd. Intrinsic line ratio for a gas with a temperature of 10000 K and a density of 100 cm$^{-3}$ is overplotted in black. The data have been binned up by two 0\farcs4 pixels. Average values in the SW and NE region, delimited with the vertical dashed lines, are overplotted in red. Error bars are displayed with respect to the
average in the SW region.  All the ratios are consistent with zero extinction except for the NE \ha/\hb\ ratio. This result suggests that
a mechanism, other than extinction by dust, is causing the \ha/\hb\ ratio to deviate from the expectations of recombination.  }
\end{figure*}

JHU/NASA sounding rocket mission 36.136~UG  was launched from White
Sands Missile Range Launch Complex-36 on 14 June 1999 at 01:40 MST. Its
purpose was to observe the hot CS of M27, provide spatial information
on the excitation state of \Htwo\ in the nebula and investigate its
dust scattering properties.  The science instrument was
a 40 cm diameter Dall--Kirkham telescope with SiC over-coated Al
mirrors, feeding a Rowland circle spectrograph with a 200\arcsec\ x
12\arcsec\ longslit, a 900 -- 1400 \AA\ bandpass and an inverse linear dispersion of $\approx$ 20 \AA\ mm$^{-1}$. The basic
configuration has been described by
\citet{McCandliss:1994,McCandliss:2000} and \citet{Burgh:2001}  and two similar
missions using this payload have flown \citep{Burgh:2002,France:2004}.
This was the first flight for a newly reconfigured spectrograph with a
holographically corrected concave grating \citep{McCandliss:2001b},
for improving the spatial resolution (3\arcsec\ spacecraft pointing
limited) and a high QE KBr coated  micro-channel plate with
double-delay line anode ($\approx$ 25 $\mu$m resolution element), similar to that used by \citet{McPhate:1999}.
 
The target was acquired at $\approx$ T+150 s, at which
time pointing control was passed to a ground based operator, who
used a live video downlink of the nebular field to make real time pointing
maneuvers.   At $\approx$ T+500 s the telescope was sealed prior to reentering the atmosphere.  The payload was recovered and post flight
calibrations were made to secure an absolute calibration.

The pointing  corrected two dimensional spectrum is shown
in the top panel of Figure~\ref{figrock}.  
\ion{C}{2} \lam 1335, \ion{C}{3} \lam 977 and \lam 1175,
\ion{He}{2}+\ion{N}{2} \lam 1085, \ion{N}{3} \dlam 989 -- 991 and
\ion{H}{1} Lyman emission lines are evident along with the continuum spectrum of the central
star.  The  bottom panel shows the extracted nebular spectrum
integrated over the length of the slit, excluding the stellar
continuum.  The strong \lya\ emission feature that fills the slit is
dominated by geocoronal emission. There is
a hint of \ion{H}{1} 2s -- 1s two photon emission starting at \lya\ and
rising towards 1400 \AA.  The dotted blue line shows the expected brightness for an column of $N(H(2s))$ = 4 $\times$ 10$^{8}$ cm$^{-2}$ \citep[e.g.][]{Nussbaumer:1984}.  There is no evidence for continuum pumped
fluorescence of \Htwo\ or dust scattered stellar continuum
down to the detector background limit of $\approx$ 5
$\times$ 10$^{-17}$~\bright\  indicated  by the red line.

\begin{figure*}[t]
\centerline{\includegraphics[]{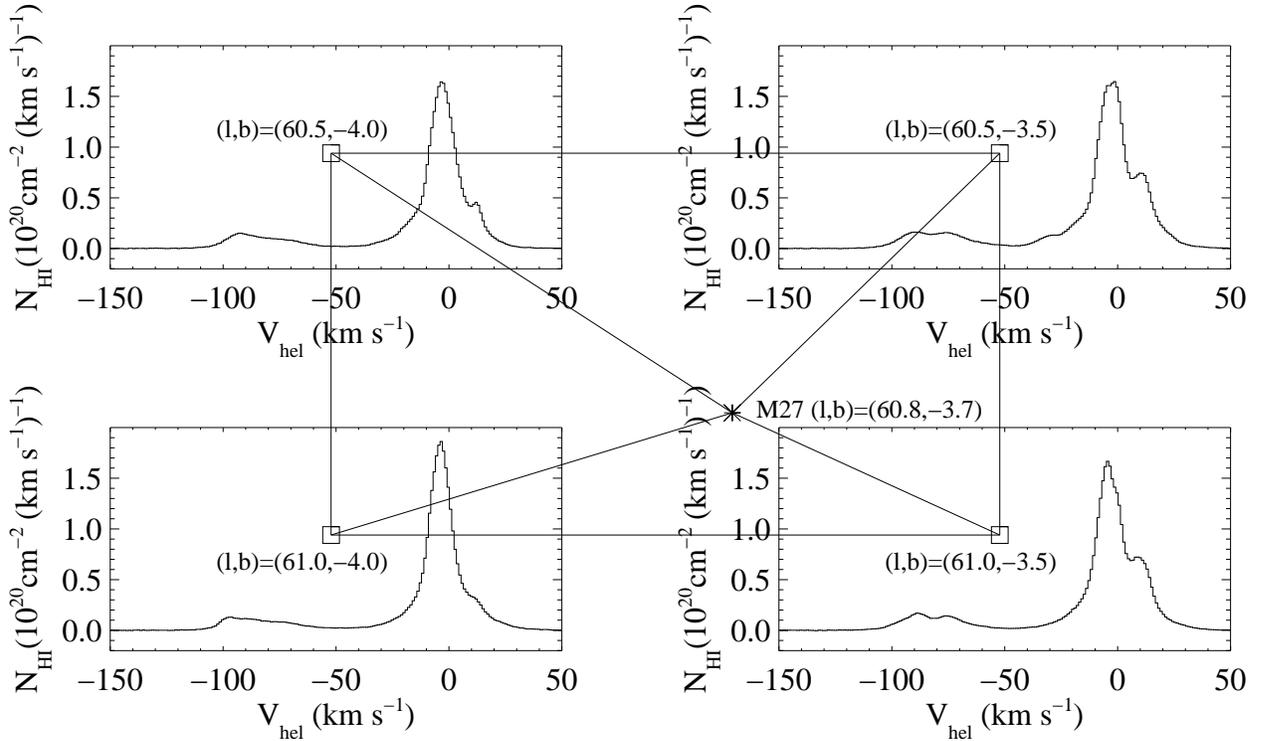}}
\figcaption[f6.eps]{ \label{dwingneib}
Nearest neighbor \ion{H}{1} 21 cm emission spectra from the Dwingloo spectral
survey atlas. These spectra were used to form a minimum spectrum and a distance-weighted mean spectrum. }
\end{figure*}

\begin{figure*}[t]
\centerline{\includegraphics[]{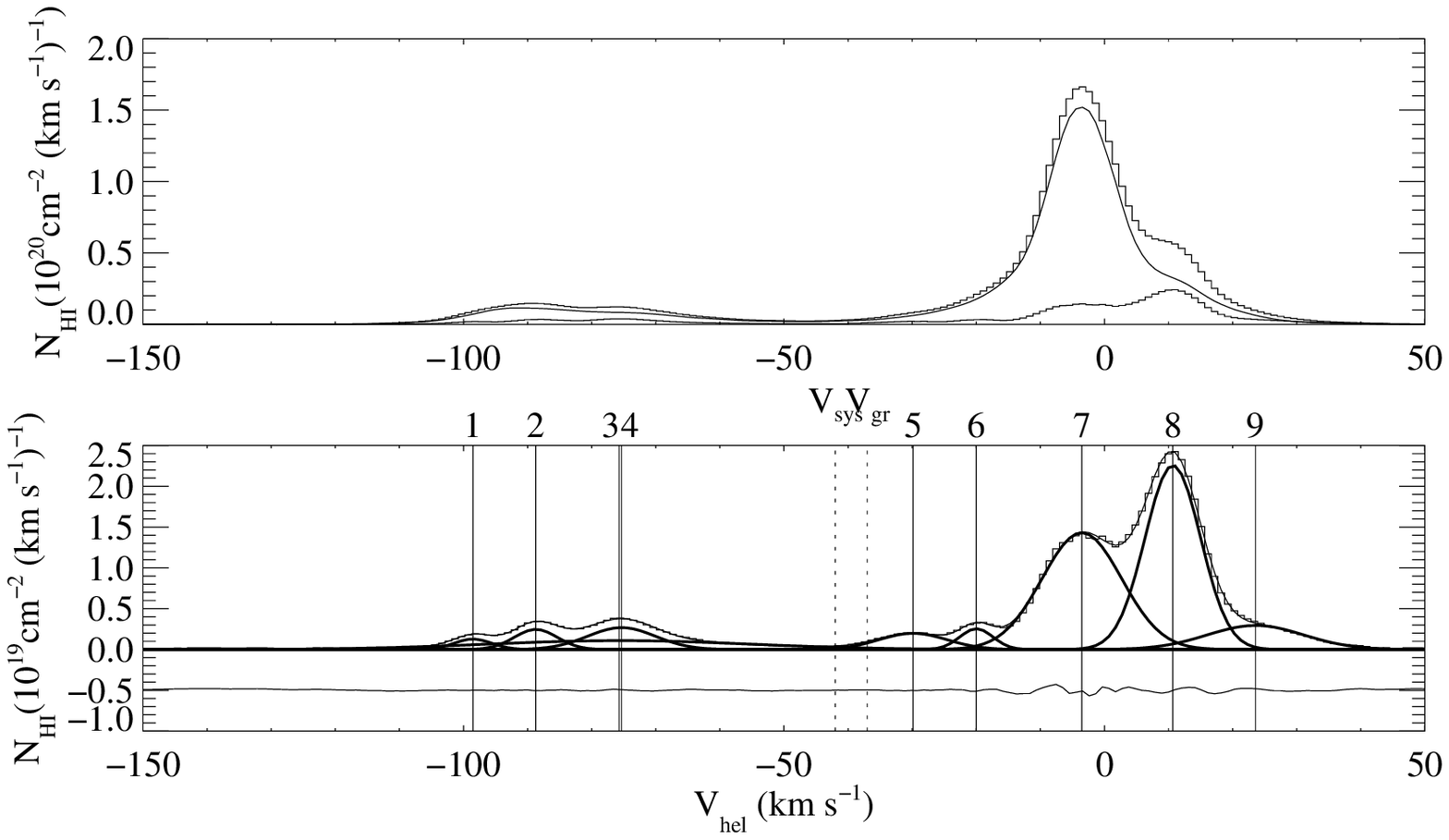}}
\figcaption[f7.eps]{ \label{dwingprofile}
Top -- Minimum spectrum is subtracted from the angular distance weighted mean spectrum to remove wide angle background emission. Vertical axis is column density per \kms. Horizontal axis is
heliocentric velocity. Bottom -- Resulting spectrum contains both foreground and some background emission.  Nine component gaussian fit is shown along with the fit residuals (offset by -5 $\times$ 10$^{18}$ cm$^{-2}$).  See Table~\ref{HIcolbvel} for total columns, velocity centroids, and doppler parameters derived from gaussian fits. }
\end{figure*}

\subsection{APO DIS Longslit Observations of Balmer Line Ratios in M27}
\label{apo}

It is curious that \citet{Hawley:1978} and \citet{Barker:1984}  both cited problems with their $H{\alpha}$/$H{\beta}$ ratios
being higher than the 2.9 ratio determined by \citet{Miller:1973}.
\citet{Hawley:1978} suggested some unidentified systematic error was
affecting the ratio at the 10\% level. \citet{Barker:1984} reached a
similar conclusion and went so far as to decrease the
$H{\alpha}$/$H{\beta}$ by 13\% before applying an extinction correction
of $c$ = 0.17. Both studies find $\approx$ 20\% variations in $H{\alpha}$/$H{\beta}$ at different locations within the nebula.

To more thoroughly investigate this phenomenon we recently acquired an extensive
series of longslit spectral scans of the nebula
with the Double Imaging Spectrograph (DIS version III) at the ARC 3.5 m
telescope at Apache Point Observatory (APO), during the nights of 1 -- 2
July 2006 commencing at 01:15 MDT.  DIS has blue and red
spectra channels with back-illuminated, 13.5 $\mu$m pixel Marconi CCDs
in a 1028 $\times$ 2048 pixel$^2$ format for recording data.  The
medium resolution gratings have inverse linear dispersions of 1.85 and
2.26 \AA~pixel$^{-1}$ respectively.  The useful spectral ranges are
3700 -- 5400 \AA\ for the blue and 5300 -- 9700 \AA\ for the red.  The
plate scales are 0\farcs42 pixel$^{-1}$ for the blue and 0\farcs40
pixel$^{-1}$ for the red. 

The hot sub-dwarf star BD +28 4211 was used as a spectrophotometric
standard \citep[][]{Bohlin:2001}. This star was observed through the large  5\arcsec\ $\times$
300\arcsec\ DIS slit for 60 seconds at an airmass of 1.18.  The M27 CS
was observed shortly thereafter for 200 seconds through the same
aperture at an airmass of 1.02.  The CS and surrounding nebula were
then observed through the narrowest 0\farcs9 $\times$ 300\arcsec\  DIS
slit for 200 seconds, with the slit held fixed on the CS. The narrowest
slit provides clean separation of the [\ion{N}{2}] \dlam 6548, 6584
lines from $H{\alpha}$. The position angle of the slit was 35\degr.
Biases, flat fields and emission line spectra were recorded at
twilight.  and the data were reduced using custom IDL code.  The
CS absolute flux is shown in Figure~\ref{sed} and will be discussed in 
\S~\ref{sedmod}.  The Balmer line longslit profiles centered on the CS, and  ratioed to  $H{\beta}$, are shown in Figure~\ref{recomb}
and will be discussed in \S~\ref{blr}.  Analysis of the full dataset from the APO run will be presented elsewhere.

\subsection{Dwingeloo \ion{H}{1} 21-cm Observation of M27}
\label{dwingeloo}

The shapes of atomic hydrogen absorption profiles are complex, as they result from ensembles of intervening ISM
absorption systems, separated in velocity space along the
line-of-sight. 
Determining the column density and doppler parameter
for individual velocity components within these broad damped and
saturated absorption profiles is  difficult without {\it a priori}
information of the line-of-slight velocity structures.  One way to gain
this information is to examine \ion{H}{1} 21 cm emission data such as found in the Atlas of Galactic Neutral Hydrogen \citep[][also known
as the Dwingeloo survey]{Hartmann:1997}.  The high spectral resolution
of the data is excellent for locating velocity components, provides an
upper limit to the amount of neutral hydrogen in the foreground, and
serves as a starting point for absorption line modeling.  The
atlas, in galactic coordinates ($l, b$)  with (0\fdg5)$^2$
cells, has a velocity spacing of 1.03 \kms.  The positions cover most
of the sky and the velocity coverage spans  --450 $\le$ v$_{lsr}$ $\le$
400 \kms\ in local-standard-of-rest coordinates.  The intensity is given in
antenna temperature per velocity (K (\kms)$^{-1}$), but can be converted to a
\ion{H}{1} column density, under the assumption of optically thin emission,
by multiplying with a conversion factor \citep[0.182 $\times$ 10$^{-19}$ cm$^{-2}$,][]{Hartmann:1997}.

The galactic coordinates of M27 are $l$ = 60\fdg84, $b$ = --3\fdg70.
The neutral portion of the nebula should be at least as large as the
H$\alpha$ halo detected by \citet{Papamastorakis:1993} ($\gtrsim$
0\fdg28) and since the Dwingeloo half power beam width is slightly
larger than the atlas grid, it is reasonable to expect some signal from
the nebula to appear in the four nearest neighbor points, i.e.  atlas
coordinates ($l$ = 61\fdg, $b$ = --3\fdg5) ($l$ = 60\fdg5, $b$ =
--3\fdg5) ($l$ = 61\fdg, $b$ = --4\fdg) and ($l$ = 60\fdg5, $b$ =
--4\fdg).  The column density profiles of \ion{H}{1} as a function of
velocity for these four grid points, in heliocentric velocity
coordinates,  are shown in
Figure~\ref{dwingneib}.

To reduce the non-local contributions we constructed a nearest neighbor
minimum spectrum under the assumption that non-local contributions are
widespread and uniform.  This assumption has been checked by examining
an area out to $\pm$1\fdg5, which showed similar structures, typified
by strong peaks near 0 \kms\ and smaller peaks near -85 \kms.  The
nearest neighbor minimum spectrum was smoothed with a five bin boxcar
average and subtracted from a similarily smoothed spectrum formed from
a distance weighted average of the same 4 nearest neighbors.  This
process is shown in Figure~\ref{dwingprofile}.  The resulting
subtraction was fit with nine Gaussian profiles spread between --99 $<
v_{hel} <$ 24 \kms.   The integrated emission column densities, rms
velocity widths (b values), and heliocentric velocities are given in
columns 2, 3, and 4 of Table~\ref{HIcolbvel}.

Column 5 shows the column densities that we adopted for the model of
the Lyman series absorptions as discussed in \S~\ref{HI}.  Absorption
line components to the blue of the systemic velocity at
--42~\kms\ occur within the expansion velocity inferred from the
optical emission lines and can plausibly be associated with the nebular
expansion.  If the absorption components to the red of --42~\kms\ are
associated with the nebula they would have to be infalling.  They are
more likely to be located in the foreground.

\subsection{STIS}
\label{stis}
The E140M spectrum of M27 (o64d07020\verb'_'x1d.fits), taken from the
Multimission Archive at Space Telescope (MAST),  was acquired for
 \HST\ Proposal 8638 (Klaus Werner -- PI).  These data are a high
level product consisting of the extracted one-dimensional arrays of flux, flux error and wavelength
for the individual echelle orders.
The data were acquired through the 0\farcs2$\times$0\farcs2
spectroscopic aperture for an exposure time of 2906 s and were reduced
with CALSTIS version 2.18.  The spectral resolution of the E140M is
given in the STIS data handbook \citep[version 7.0][]{Quijano:2003} as
$R$ $\approx$ 48,000 (${\delta}V$~=~6.25~\kms).  The intrinsic line
profile for the 0\farcs2$\times$0\farcs2  aperture has a gaussian core
with this width, but there is non-negligible power in the wings.
Comparison of the \FUSE\ and STIS wavelength scales for the CS revealed a
discrepant offset.  The establishment of a self consistent wavelength
scale is discussed in \S~\ref{profiles}.

\section{Data Analysis}

Our primary purpose is to quantify the excitation state of the
\Htwo\ in the diffuse nebular medium.  The column densities
of the individual ro-vibrational states were determined using
curve-of-growth analysis.  We detect essentially imperceptible
reddening of the stellar SED over a decade of wavelength coverage, in
contrast to the reddening determinations using the Balmer decrement
method as reported in the literature (\S~\ref{previous}).   We show new
high precision longslit Balmer decrement ratios (H$\alpha$/H$\beta$,
H$\gamma$/H$\beta$, and H$\delta$/H$\beta$) where H$\alpha$/H$\beta$
appears to be reddened in certain regions of the nebula, but the higher
order decrements appear unreddened everywhere.  A potential resolution of the
reddening conundrum will be discussed in \S~\ref{disrecomb}.

We also produce an estimate of the total atomic hydrogen  column
density in the nebula to enable a discussion of  \Htwo\ formation and
destruction processes.  An upper limit is placed on the column density
of CO in the diffuse medium and an analysis of the excitation of
\ion{C}{1} fine structure lines will provide a diagnostic of nebular
pressure  and insight into the \Htwo\ exitation mechanism.
Absorption line profiles of high and low ionization species
compared to the neutral profiles reveal the ionization
stratification of the outflow and provide a velocity description of the nebular
kinematics.

\begin{figure*}[t]
\centerline{\includegraphics[]{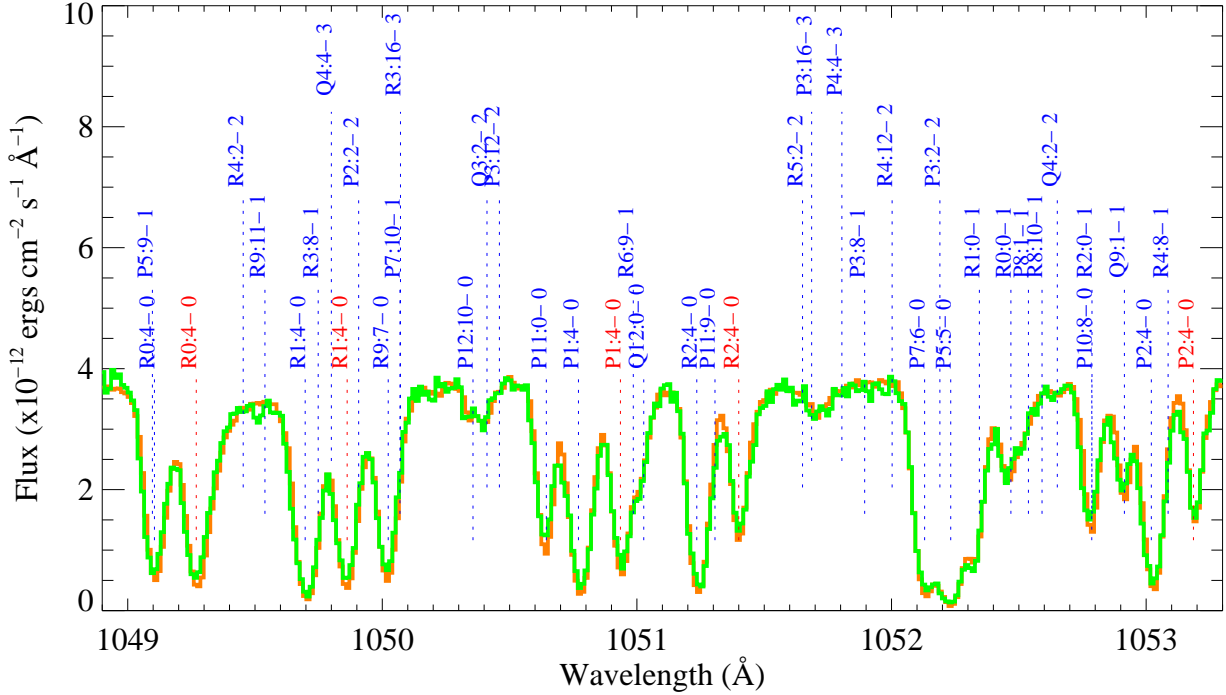}}
\figcaption[f8.eps]{ \label{blueandred} 
Region near the \Htwo\ (4--0) band identifying the
high excitation velocity component at --33 \kms\ (blue) and the low excitation component at +14 \kms\ (red) as measured in the CS restframe.  \FUSE\ s12 and s21 spectra are overplotted in orange and green respectively.}
\end{figure*}

\begin{figure*}[t]
\centerline{\includegraphics[]{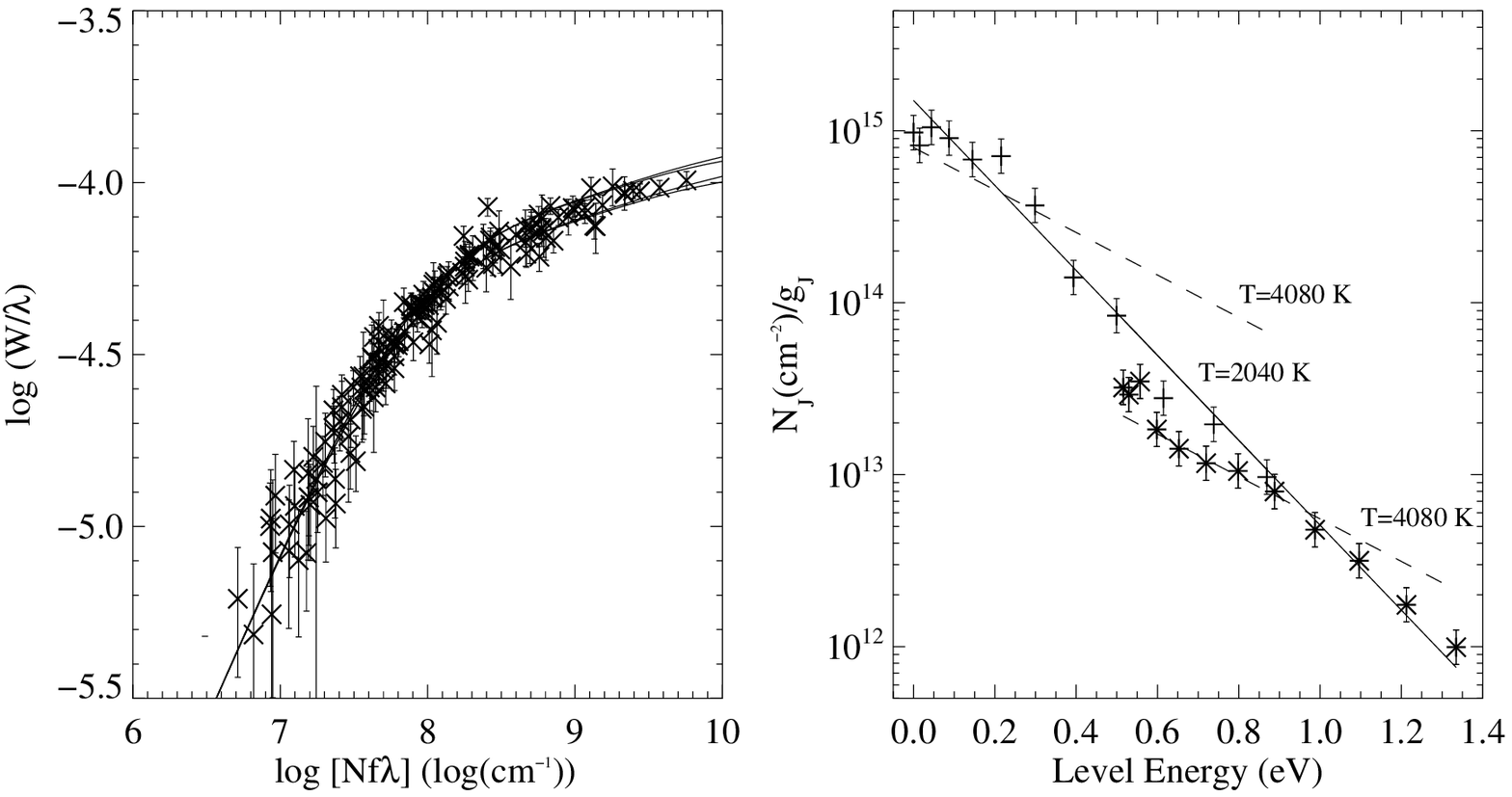}}
\figcaption[f9.eps]{\label{ncolcognew}
Upper -- curves-of-growth for doppler parameters of 6 and
7~\kms\ along with the data.  A doppler parameter of 6.5~\kms\ was
adopted. Lower -- column densities, normalized by level
statistical weight, as a function of the level energy.    The \vpp\ =
0, 0 $\le J^{\prime\prime} \le $11 columns are shown with + marks, and the \vpp\ = 1, 0
$\le J^{\prime\prime} \le $11 columns are shown with * marks.  Overplotted (solid line) is
a single temperature Boltzmann distribution for $T$ =2040 K.  There are
statically significant deviations from the single temperature
population, although the higher 7 $\le J^{\prime\prime} \le $ 11 states are in
agreement.  The lower 0 $\le J^{\prime\prime} \le $ 6 states appear to have a
flatter slope indicative of a temperature a factor of two higher as the
dashed lines indicate.
}
\end{figure*}

\subsection{Curve-of-Growth for the Molecular Hydrogen Absorption Components}
\label{cogsec}

It seems with a first glance at Figure~\ref{edge} that determining the
column densities for the ro-vibrational transitions and identifying
velocity components will be daunting.  However, the regular spacing of
\Htwo\ lines arising from common rotational states, along with the
monotonic variation in transition strength with increasing upper vibrational
level, makes identification of discrete velocity components straightforward.  The high line density becomes an advantage, virtually guaranteeing  a few lines for a given rotational state will be
unblended.  This allows column densities to be determined by a straight forward
measurement of the equivalent widths for use in a curve-of-growth analysis.

We have identified two distinct components with heliocentric velocities
of --75  and --28   $\pm$ 2 \kms.  The blueward velocity component is
the  more highly excited, allowing the construction of curves-of-growth
for all the ro-vibrational levels with \Jpp\ $\le$ 11 and \vpp $\le$ 1
of the ground electronic state \xsig.\footnote{For rotational and
vibrational quantum numbers in a transition leading to an upper
electronic state (e.g. either \bsig\ or \cpi) from the ground state
(\xsig) the convention is to designate the upper state with a single
prime (\Jp, \Vp) and the lower state with a double prime (\Jpp, \vpp).}
The redward component has no extraordinary excitation with
measurable absorption only found for \Jpp\  $\le$ 3 in  \vpp\ = 0.
Figure~\ref{blueandred} shows a small region around the (\Vp--\vpp) =
(4--0) band near 1050\AA\ where the ``blue'' and ``red'' velocity
components are identified.  Blueward absorptions for (\Vp--1), (\Vp--2)
and (\Vp--3) in this region are also displayed.

A semi-autonomous method was developed to measure the equivalent widths
of the lines, based in part on the completely autonomous method used by
\citet{McCandliss:2001a}.  All the lines for a given velocity offset
and rotational state were located within a summation interval initially
13 -- 15 pixels wide.  The pixels adjacent to either side of the
summation interval were used to define continuum points, through which
a straight line was fit to serve as the model for the continuum in the
summation region.  The degree of blending was assessed interactively by
plotting the spectrum surrounding each individual line in a (-150, 50)
\kms\ interval and overplotting all the known atomic and molecular
features.  Lines with evidence of blending were
rejected from the curve-of-growth.  Examination of the lines for blends
provided an opportunity to fine-tune the summation interval and
continuum placement.  The equivalent width measurements were carried
out on both the s12 and s21 spectra (see \S~\ref{fusespec}), using the
same summation and continuum intervals.  Allowance was made for the
slight local mismatches in velocity scale that exist between the s12
and s21 spectra, by locating the minimum within the
integration region and adjusting the interval accordingly.

Initial curves-of-growth were constructed from the equivalent width
measurements using the wavelengths and oscillator strengths from
\citet{Abgrall:1993a, Abgrall:1993b}. We required at least 2 unblended
lines for these curves.  Lines from transitions with \Jpp\  $\ge$ 11
and  \vpp\ $\ge$ 2 are detected, but they are weak and too few to
construct a reliable curve-of-growth.  Independent $\chi^2$ fittings
were performed for each curve by varying the doppler parameter over 2
-- 10 \kms\ and the logarithm of the column density (in cm$^2$) over 13
-- 18 dex.  The doppler parameters thus derived varied from 4 to 8
\kms\ and the logarithmic column densities ranged over $\approx$ 13.5
-- 17 dex.

\begin{deluxetable}{ccccc}
\tablecolumns{5}
\tablewidth{0pc}
\tablecaption{Derived Nebular H$_2$ \xsig\ Column Densities\tablenotemark{*} \label{h2col}}
\tablehead{ \colhead{J$^{\prime\prime}$} &\colhead{$\log{N(v^{\prime\prime}=0)}$} & \colhead{n$_{cog}$} & \colhead{$\log{N(v^{\prime\prime}=1)}$}  & \colhead{n$_{cog}$} \\
& \colhead{$\log$(cm$^{-2}$)}& \colhead{lines} & \colhead{$\log$(cm$^{-2}$)}& \colhead{lines} }
\startdata
      0&  15.0   &   2	&13.6	&   3	\\
      1&  15.9   &   11	&14.4	&   5	\\
      2&  15.7   &   8	&14.2	&   3	\\
      3&  16.3   &   4	&14.6	&   9	\\
      4&  15.8   &   4	&14.1	&   10	\\
      5&  16.4   &   8	&14.6	&   12	\\
      6&  15.7   &   8	&14.1	&   5	\\
      7&  15.8   &   5	&14.5	&   11	\\
      8&  15.2   &   7	&13.9	&   2	\\
      9&  15.2   &   11	&14.3	&   4	\\
     10&  14.6   &   13	&13.5	&   2	\\
     11&  14.8   &   5	&13.8	&   2	\\
\enddata
\tablenotetext{*}{Systematic error =
$\pm$ 0.1 dex. \\The doppler parameter is 6.5 $\pm$ 0.5 \kms.}
\end{deluxetable}

Comparing results  from the  s12 and s21 measurements revealed a few
disagreements for the doppler parameter (and thus column density)
derived for the same rotational state.  This produced an inconsistent
modulation in the ortho-para ratio, expected to be $\approx$ 3:1 for
the odd to even rotational states.  The inconsistencies were traced to
curves-of-growth with the most saturated lines and lowest number of
points. Restricting the allowed doppler parameters in the fit to either
6 or 7 \kms\ (the most common values) resolved the column density
discrepancy and resulted in consistent ortho-para modulation. It
also reduced the scatter when all the curves-of-growth for the
individual rotational states from the s12 and s21 spectral measurements
were combined (top panel of  Figure~\ref{ncolcognew}).  The column
densities for the individual ro-vibrational levels are listed in
Table~\ref{h2col} along with the number of lines used for each
curve-of-growth.  The errors are dominated by placement of the
continuum.  Experiments were performed to gauge the effect of
systematic continuum offsets on the derived column densities.  We found
the column densities were consistent to within $\pm$ 0.1 in
the dex for offsets within the local continuum signal-to-noise.

The total column density for the hot \Htwo\ component at
--75 \kms\ is N(H$_2$)=7.9  $\pm$ 2 $\times$ 10$^{16}$ cm$^2$ with a
doppler velocity of 6.5 $\pm$ 0.5 \kms.  Curve-of-growth
analysis on the cold non-nebular component, for which we wish only to identify the strength of its aborption features,  yielded
a total column density of 1.3 $\pm$ 5 $\times$ 10$^{17}$ cm$^2$ with a
doppler velocity of  5 \kms.  The rotational distribution is well
approximated with a temperature of 200 K.

In the bottom panel of Figure~\ref{ncolcognew} we show the population
density of the ro-vibrational levels as a function of level energy.  We use (+) marks to indicate the columns for \Jpp\
in the \vpp\ = 0 state and (*) marks to indicate the columns for \Jpp\ in the \vpp\ = 1 state.  The straight solid line is the
best fit single temperature model ($T$ = 2040 K), assuming a Boltzmann
distribution of level populations.  This model is not successful, as
there are statistically significant deviations from a single
temperature Boltzmann destribution.

The variation of population density as a function of level energy is
quite different from what is usually observed in the cold ISM.
Typically the first two rotational levels are consistent with a rather
steep slope with a temperature $T_{01} \sim$ 80 K while the higher
rotational levels flatten, giving the appearance of a higher
``temperature'' of several hundred degrees or more
\citep{Spitzer:1973}.  The excess column in the $J >$ 1 levels has long
been thought to be caused by far-ultraviolet continuum fluoresence, but
recently \citet{Gry:2002} has questioned this hypothesis. They favor 
a true two temperature model for the ISM.  

Regardless, here we have the opposite case.  In each vibrational level
\vpp\ = 0, 1 the trend is for the lower rotational states to have
flatter slopes (higher temperature) than the higher rotational states.
Reproduction of the energy level distribution found here will be an
interesting challenge for \Htwo\ excitation models
\citep[c.f.][]{Spitzer:1974, Black:1976, Shull:1978, vanDishoeck:1986,
Sternberg:1989b, Draine:1996}.  We note how the modeling by
\citet{Spitzer:1974} shows qualitatively how the ``curvature'' of the
population density can go from ``concave'' to ``convex'' as the
parameters  change from a low temperture, density and photoexcitation
rate environment to either a high photoexcitation rate or a high
density environment.  However, they do not attempt to explore
temperatures in excess of 1000 K.

\subsection{Central Star Spectral Energy Distribution and Extinction}
\label{sedmod}

SED models with CS parameters taken from the literature provide a
fairly precise  match to the stellar continuum observed by  \FUSE.
This allows us to determine the line-of-sight extinction, and provides
a reasonable stellar continuum to use for assessing the success of our
model of atomic and \Htwo\ absorption.  The combined SED and
absorption  model has been used by \citet{McCandliss:2007} to identify
63  photospheric, nebular and non-nebular absorption features of
ionized and neutral metals, lurking amid the sea of hydrogen features.

In Figure~\ref{sed} we show a log-log plot of the CS SED from 900 --
20000 \AA, as measured using \FUSE, \IUE, and the APO DIS, along with
the optical photometry
\citep{Tylenda:1991,Ciardullo:1999,Benedict:2003}.  Over-plotted is a
synthetic stellar flux interpolated from the grid of \citet{Rauch:2003}
with $\log{g}$ = 6.5, T = 120,000 K, and ratio of H/He = 10/3 by mass.
This model includes no metals and is consistent with, although slightly
hotter than, the quantitative spectroscopy of \citet{Napiwotzki:1999}.
It is slightly cooler than the determination of \citet{Traulsen:2005}.
The temperature and gravity adopted for our model has slightly less
pressure broadening and is a better match to the observed Lyman lines
towards the series limit.  We also adopt a stellar mass of 0.56
$M_{\sun}$ as suggested by post asymptotic giant branch evolutionary
tracks.

Use of this mass along with the above gravity required a distance of
466 pc to match the absolute flux, which is at the upper limit given by
\citet{Benedict:2003}.  Beyond our immediate need to match the absolute
flux for the given gravity and mass there is no particular reason to
prefer this distance over that derived by \citeauthor{Benedict:2003}.
Questions regarding the acceptable uncertainty in distance, absolute
flux and derived stellar parameters are best left for a stellar model
specifically tailored to include the effects of metals, gravity and
evolutionary state.  For our purpose, the high stellar temperature
places the SED longward of the Lyman limit in the Rayleigh--Jeans
regime, so the shape of the SED is insensitive to our assumptions of
temperature, gravity, metallicity, mass and distance.  Consequently,
long-range changes in SED shape produced by reddening can be
constrained with a high degree of confidence.

The model flux is shown in Figure~\ref{sed}  with reddenings  of E(\bv)
= 0.00, 0.005 and 0.05 assuming the R$_V$ = 3.1 curve of
\citet{Cardelli:1989}.   An extinction of E(\bv) =  0.05 is clearly too
high.  As a convenience we adopt a very modest extinction of E(\bv) =
0.005, as seen most clearly in Figure~\ref{edge} to make the continuum
towards the Lyman edge match the stellar SED.  The lack of any
significant extinction is roughly consistent with the total nebular
hydrogen,  derived in \S~\ref{HI}, and the standard conversion for
color excess to total hydrogen column of $N(H)/E(\bv)$ = 5.8 $\times$
10$^{21}$ cm$^{-2}$ mag $^{-1}$ \citep{Bohlin:1978}.

\subsubsection{Balmer Line Reddening}
\label{blr}

Reddening by dust in PNe is often estimated by comparing the measured
Balmer emission ratios to the intrinsic ratios produced by hydrogen
recombination. The intrinsic intensity ratio (in flux units) of
$I(H\alpha)$/$I(H\beta)$ $\simeq$ 2.859 at an electron temperature of
10$^4$K and density of 10$^2$ cm$^{-3}$, is accurate to within $\approx
\pm$ 5\% for a wide range of the electron temperature and density,
(5000 $< T_{e}$(K) $<$ 20000,  10$^2 < N_e($cm$^{-3}) <$ 10$^6$)
\citep{Brocklehurst:1971}. The extinction parameter $c$ is given in
\begin{equation} I(\lambda)/I(H\beta) = F(\lambda)/F(H\beta)
10^{cf(\lambda)} \end{equation} where the intrinsic ratio is
$I(\lambda)/I(H{\beta})$, the observed  ratio is
$F(\lambda)/F(H{\beta})$ and $f(\lambda)$ is the line attenuation,
relative to \hb, at any wavelength for the given extinction curve
\citep[c.f.][]{Miller:1972, Cahn:1976}.  The extinction parameter may
be converted to selective extinction by adopting the ratio of
$c/E(\bv)$ = 1.5 as shown by \citet{Ciardullo:1999} in their Figure 5.

Balmer emission line spatial profiles for \ha, \hb, \hg, and \hd\ in a
$\pm$ 140\arcsec\ region to either side the CS at a position angle of
35 $\degr$, were acquired as discussed in \S~\ref{apo}.  The line
ratios,  with respect to $H\beta$, are displayed in Figure~\ref{recomb}
with a two pixel (0\farcs9) binning.  The reddenings derived from these
ratios (Figure~\ref{recomb}) are inconsistent and are summarized in
Table~\ref{balrat}.  The first column specifies the line ratio, the
second is the extinction multipler taken from \citet{Barker:1984}, the
third is the intrinsic line ratio for a temperature of 10,000 K and a
density of 100 cm$^{-3}$ taken from \citet{Brocklehurst:1971}, the
fourth and fifth columns are the average line ratios and standard
deviations for the SW and NE regions after rebinning the data by
sixteen pixels (7\farcs2), and the sixth and seventh columns are the
derived selective extinctions and associated errors.  The NE \ha/\hb\
ratio suggests $E(\bv)$ = 0.10 $\pm$ 0.02, while all the rest of the
ratios are consistent with much lower or no extinction.

The line ratio averages in the fiduical SW and NE regions have standard
deviations of 1.5 -- 3~\% after rebinning by sixteen pixels.  The large
error bars in the derived $E(\bv)$ serve to emphasize that Balmer line
ratio determinations to a precision of better than 1\% are required for
$E(\bv) <$ 0.01.  We note that in the immediate vicinity of the star
\ha/\hb\ = 3.03  yielding   $E(\bv) $ = 0.051, which is much too large to
be consistent with the observed far ultraviolet SED (see \S~\ref{sedmod}), assuming the line-of-sight extinction is close to the Galactic standard.  

We concluded in the previous section that dust along the diffuse line
of sight produces $E(\bv) \le $ 0.01 mag.  Here we find the
extinction in the extended medium is low to nonexistent.  We conclude that dust is not widespread, although it may be confined to the clumpy medium.  The
inconsistancies in Balmer reddening measures suggest that a process other than
extinction by dust is contributing to \ha/\hb\ ratio excess. We will discuss this issue further in \S~\ref{disrecomb}.

\begin{figure*}
\figurenum{10a}
\centerline{\includegraphics[height=8.5in]{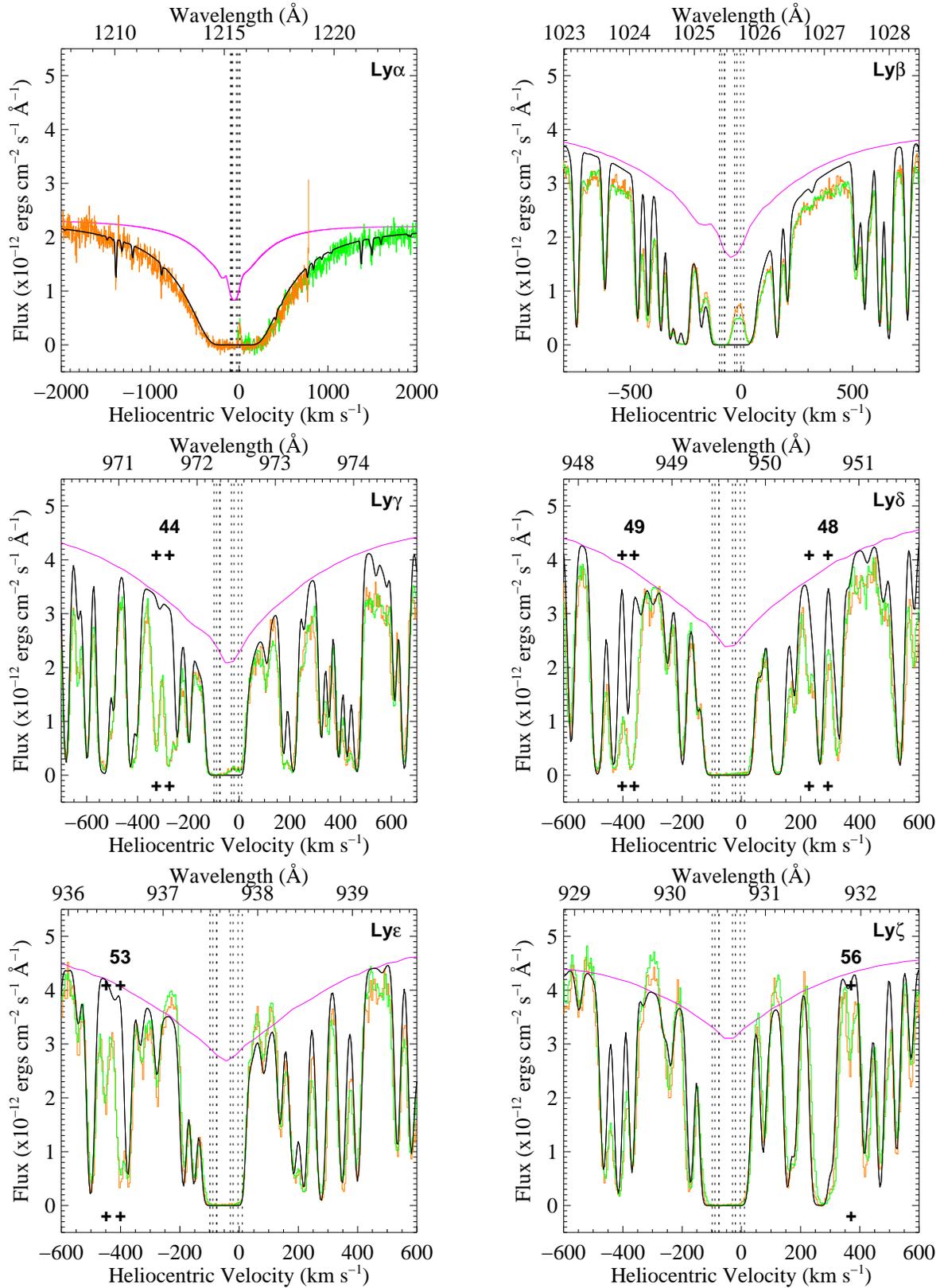}}
\figcaption[f10a.eps and f10b.eps]{ \label{hiproflong}
Lyman series model fits (Ly$\alpha$ -- $\mu$). The long dashed lines near the center of the profiles mark the velocities of the absorption components determined from the Dwingeloo data.  In Ly$\alpha$ the two STIS E140M 
orders are shown in orange and green.  The \FUSE\ data are shown
as orange (s12) and green (s12). The stellar continuum is in purple and the absorption model is in black.  The numbered +'s mark absorption features tabulated in Table~2 of \citet{McCandliss:2007}.}
\end{figure*}

\begin{figure*}
\figurenum{10b}
\centerline{\includegraphics[height=8.5in]{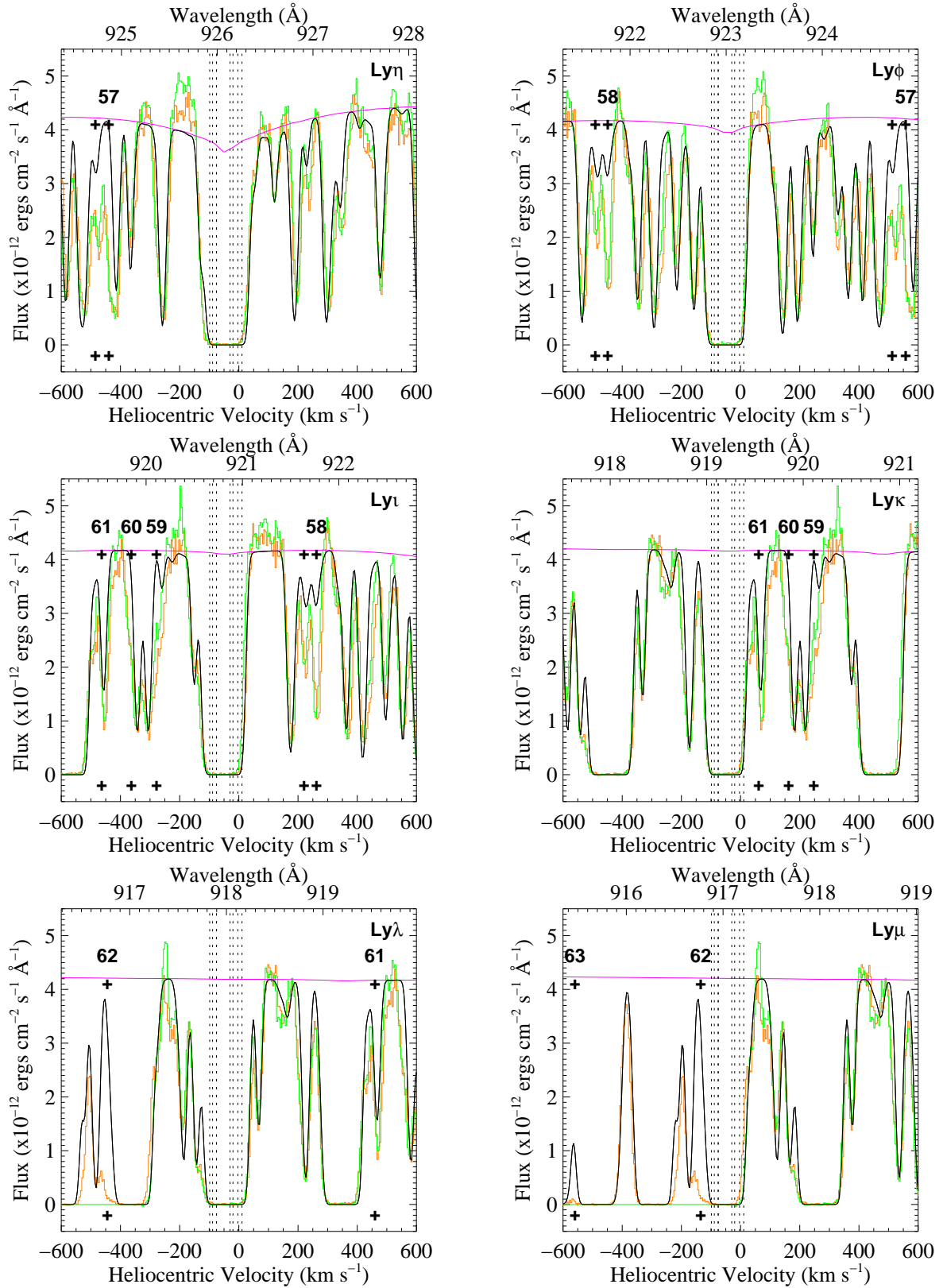}}
\caption{ 
Lyman series model fits (Ly$\eta$ -- $\mu$). Colors and marks are the same as in Figure~\ref{hiproflong}.}
\end{figure*}

\setcounter{figure}{10}

\subsection{Atomic Hydrogen Absorption Model}
\label{HI}

\begin{deluxetable}{ccrrc}
\tablecolumns{5}
\tablewidth{0pc}
\tablecaption{\ion{H}{1} 21 cm  Velocity Component Column Densities \tablenotemark{*}\label{HIcolbvel}}
\tablehead{\colhead{Component}& \colhead{$\log{N_{em}}$}\tablenotemark{$\dagger$} & \colhead{$b$} & \colhead{$V_{hel}$} & \colhead{$\log{N_{abs}}$}\tablenotemark{$\ddagger$}\\ & \colhead{$\log$(cm$^{-2}$)} & \colhead{(km s$^{-1}$)} & \colhead{(km s$^{-1}$)} & \colhead{$\log$(cm$^{-2}$)}}
\startdata
1&19.0& 2.9&-98.5& 18.9 \\
2&19.3& 3.4&-88.7& 19.0 \\
3&19.6& 5.3&-75.3& 19.0 \\
4&19.7&20.4&-75.7& 18.2 \\
5&19.4& 5.3&-29.9& 19.4 \\
6&19.2& 2.6&-20.0& 19.2 \\
7&20.3& 6.2& -3.5& 19.3 \\
8&20.4& 4.3& 10.7& 18.0 \\
9&19.7& 7.5& 23.6& \nodata \\
\enddata
\tablenotetext{*}{See figure~\ref{dwingprofile}.}
\tablenotetext{$\dagger$}{Emission columns are upper limits. Error is $\pm$ 0.15 dex.}
\tablenotetext{$\ddagger$}{Absorption columns used in \ion{H}{1} model.}
\end{deluxetable}

\begin{deluxetable*}{ccccccc}
\tablecolumns{7}
\tablewidth{0pc}
\tablecaption{Extinctions Derived from Balmer Ratios \tablenotemark{*} \label{balrat}}
\tablehead{ 
\colhead{}& 
\colhead{$f(\lambda)$\tablenotemark{$\diamond$}} & 
\colhead{Int Ratio\tablenotemark{$\dagger$}} & 
\colhead{Obs Ratio SW\tablenotemark{$\ddagger$} } & 
\colhead{ Obs Ratio NE\tablenotemark{$\ddagger$}} & 
\colhead{ $E(\bv)$ SW}& 
\colhead{ $E(\bv)$ NE}  
}
\startdata
$\frac{H\alpha}{H\beta}$ &-0.33	&2.859	&2.868 $\pm$ 0.041  &3.205 $\pm$ 0.066	&0.003 $\pm$ 0.013  &0.100 $\pm$ 0.018	\\
$\frac{H\gamma}{H\beta}$ &0.15	&0.4685	&0.4627 $\pm$ 0.0093&0.4724$\pm$ 0.0144	&0.024 $\pm$ 0.039  &-0.016$\pm$ 0.059	\\
$\frac{H\delta}{H\beta}$ &0.20	&0.2591	&0.2585 $\pm$ 0.0075& 0.2508$\pm$ 0.0082	&0.003 $\pm$ 0.042  &0.047 $\pm$ 0.048	\\
\enddata
\tablenotetext{*}{Data rebinned by 16 pixels}
\tablenotetext{$\diamond$}{$f(\lambda)$ from \citet{Barker:1984}.}
\tablenotetext{$\dagger$}{Instrinsic line ratios from \citet{Brocklehurst:1971}.}
\tablenotetext{$\ddagger$}{SW values are averaged over (--80,--50)\arcsec, the NE  over (50,80)\arcsec; see dashed lines in Figure~\ref{recomb}. } 
\end{deluxetable*}

In Table~\ref{HIcolbvel} we list the column densities, doppler
velocities and heliocentric velocity offsets inferred from the
Dwingeloo data (\S~\ref{dwingeloo}).  The 21-cm emission column
densities listed in column 2 have contributions from both background
and foreground emission, so they provide an upper limit to the expected
foreground absorption column densities.  We are mostly concerned with finding
reasonable numbers for the nebular absorption, i.e. those components
shortward of the stellar systemic velocity of --42 \kms; components 1
-- 4.  We will refer to components 5 -- 9 as the non-nebular ISM
components.

The adopted neutral hydrogen absorption model, the components of which
are listed in column 5,  was determined with the following procedure.
We initially allowed  the two components nearest to $V_{sys}$ (5 and 6)
to retain the maximum column suggested by the emission line fit.  The
two components nicely filled the centers of the Ly$\alpha$ -- Ly$\mu$
lines.  We then moved to the blueward side and reduced the column of
component 1 until a good agreement was found in the most saturated
lines Ly$\eta$ -- Ly$\mu$.  We continued to add in components 2, 3 and 4
reducing the columns as necessary to smoothly match the absorption in
the saturated lines. Once we were satisfied with the blue edge fit we
proceeded to the redward components.  It was determined that component 9
produced too much red edge absorption in all the Lyman series lines and
it was eliminated from further consideration. We found that only a
modest amount of component 8 was necessary to match the redward edge of
Ly$\alpha$.  Finally component 7 was added to reduce the absorption
gap between components 6 at --20 \kms\ and 8 at 11 \kms.  Changing the column density of any
one component by the error in the emission line fit (0.15 dex) did not
appreciably cause deviations of the resulting model with the
data.   The constraints on the formation and destruction of molecular
hydrogen within the nebula, which we discuss in \S~\ref{fordest}, are
immune to the details of the velocity distribution of \ion{H}{1}
derived from the Dwingeloo data. 
\begin{figure*}[t]
\centerline{\includegraphics[]{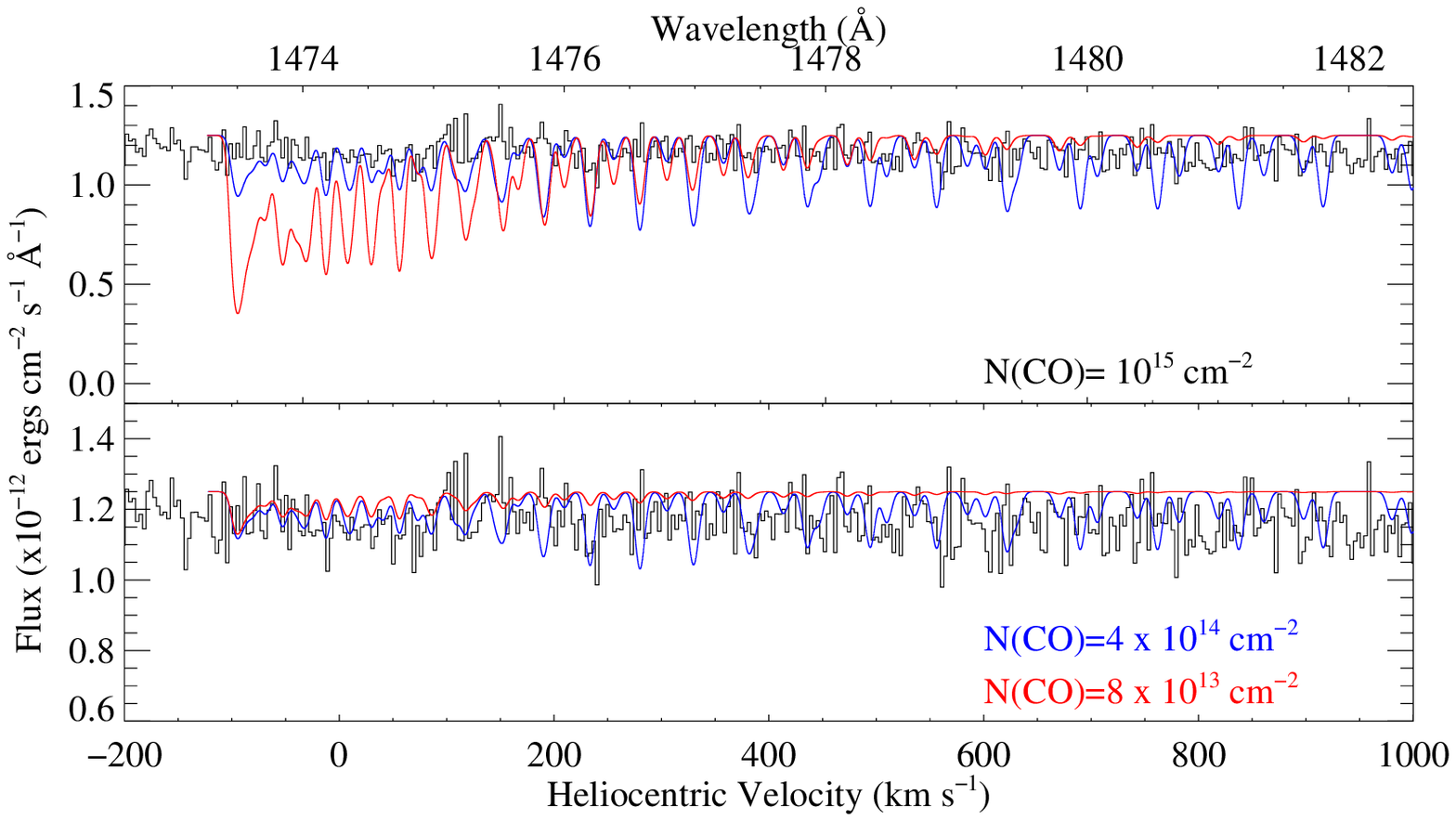}}
\figcaption[f11.eps]{ \label{co}
\HST\/STIS E140M spectrum of the region around CO A--X (2--0), the
strongest band in the A--X absorption series.  Top -- Two different excitation temperature models (3000 K, blue) and (300 K,red) of the (2--0) band using
the same column, 10$^{15}$ cm$^{-2}$, as suggested from CO mm
measurements.  Bottom -- noise bounded upper limits to the
line-of-sight CO column for the two excitation temperatures.  Note the
scale change between the two panels.  The upper limit drops for
decreasing excitation temperatures, because the population becomes
concentrated in the lower rotational levels as temperature decreases.
}
\end{figure*}

Figure \ref{hiproflong} shows an overlay of the
adopted atomic and \Htwo\ absorption model on the
Ly$\alpha$ -- Ly$\mu$ lines from STIS and \FUSE.  The Ly$\alpha$
profile from the STIS E140M observation spans two orders, plotted as
orange and green.  The rest of the Lyman series are from \FUSE\ with
the s12 spectrum  shown in orange and s21 shown in green.  The stellar
SED is shown in purple and the model is in black.  The locations of the
the individual \ion{H}{1} velocity components from
Table~\ref{HIcolbvel} are marked with vertical dotted lines.  The model includes the \Htwo\ absorption determined in
\S~\ref{cogsec}.  A comparison of the hydrogen absorption model against
the s12 and s21 spectra for the entire \FUSE\ wavelength range along
with metallic absorption system  identifications, arising from the
photosphere, the nebula and non-nebular ism, can be found in
\citet{McCandliss:2007}.  We see that the model continuum  in the vicinity of Ly$\beta$ is $\sim$ 10\% too high,
while at the shortest wavelengths it is low by a similar amount.  There
are also indications of absorption from \ion{O}{1} and \ion{N}{2}
features \citep[Table 2]{McCandliss:2007}.

Although the agreement of the model with the data is
excellent, the procedure adopted here for constraining the neutral
hydrogen column in the  nebula is not ideal and in all likelihood
provides a non-unique solution.  However, the result is at least
plausible, well within factors of two, given the uncertainty associated with the large background
subtraction necessitated by the coarse angular resolution of the
Dwingeloo data cube.  A high angular resolution 21 cm mapping of the
nebula would provide a more useful constraint on the velocity
components, column densities and doppler widths associated with the
nebula.  We are mainly concerned with the total \ion{H}{1} column density
of the nebula rather than the individual components.
The sum of components 1 -- 4 is N(\ion{H}{1})$_{neb}$ = 3.0  $\pm$ 1 $\times$ 10$^{19}$ cm$^2$.  The non-nebular  components 5 -- 8 total to N(\ion{H}{1})$_{non}$ = 6.2  $\pm$ 1 $\times$ 10$^{19}$ cm$^2$.

\subsection{CO Upper Limit and the \ion{C}{1} Pressure Diagnostic} 

\begin{figure*}[t] \vspace*{.25in}
\centerline{ \includegraphics[]{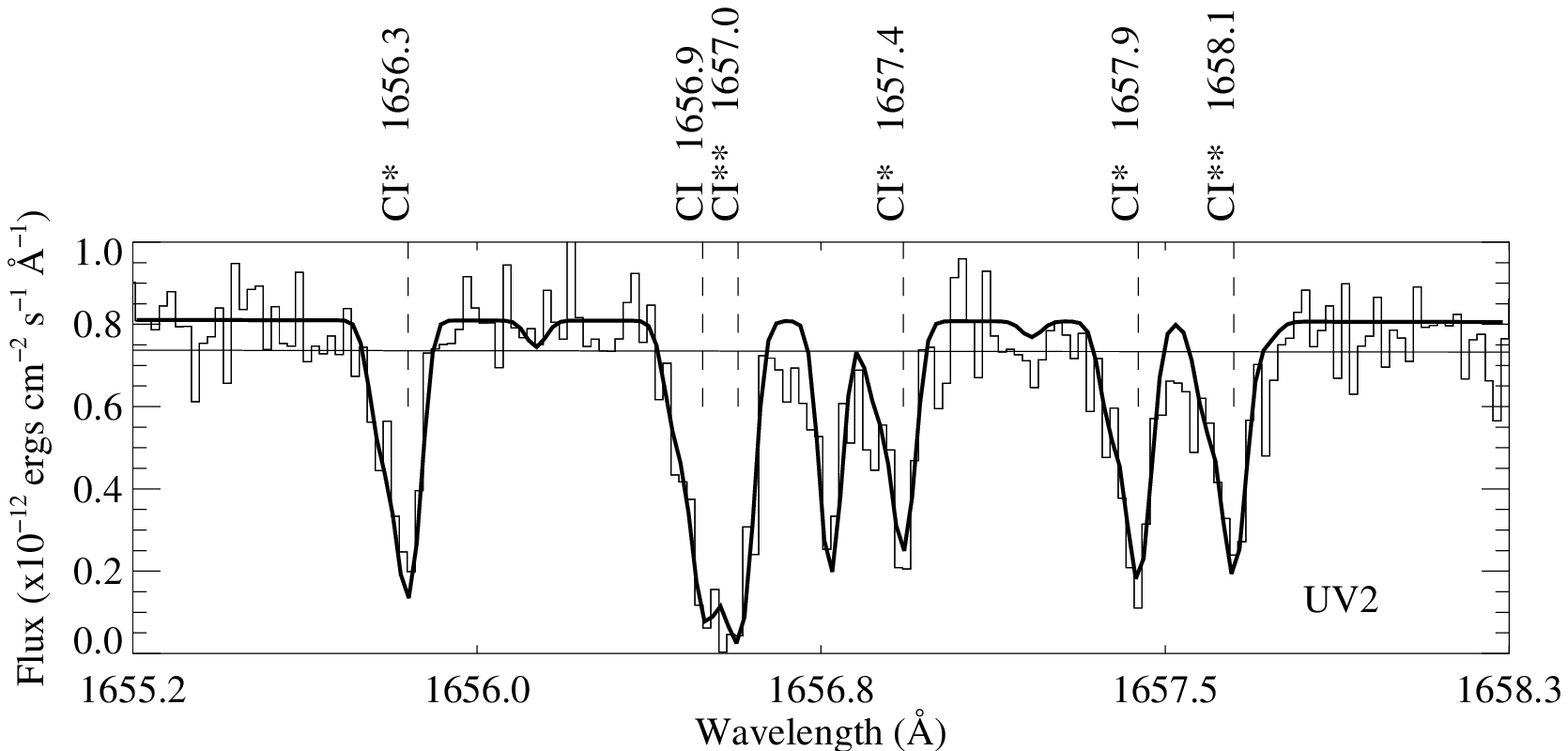}}
\figcaption[f12.eps]{ \label{cifit}
Model of the neutral carbon multiplet  UV2 (smooth line) is overplotted
on the observed \HST\/STIS spectrum (histogram).  The nebular \ion{C}{1}, \ion{C}{1}* and \ion{C}{1}** lines are labeled.  The cold non-nebular \ion{C}{1}, \ion{C}{1}*
are not labeled but the absorption is evident.  There is no non-nebular \ion{C}{1}** component. The model SED continuum is the lower smooth line.  It was increased by 10\% to improve the fit.  The $f1$ and $f2$ fractions are relatively insensitive to continuum adjustments. }
\end{figure*}

\begin{figure*}
\centerline{\includegraphics[]{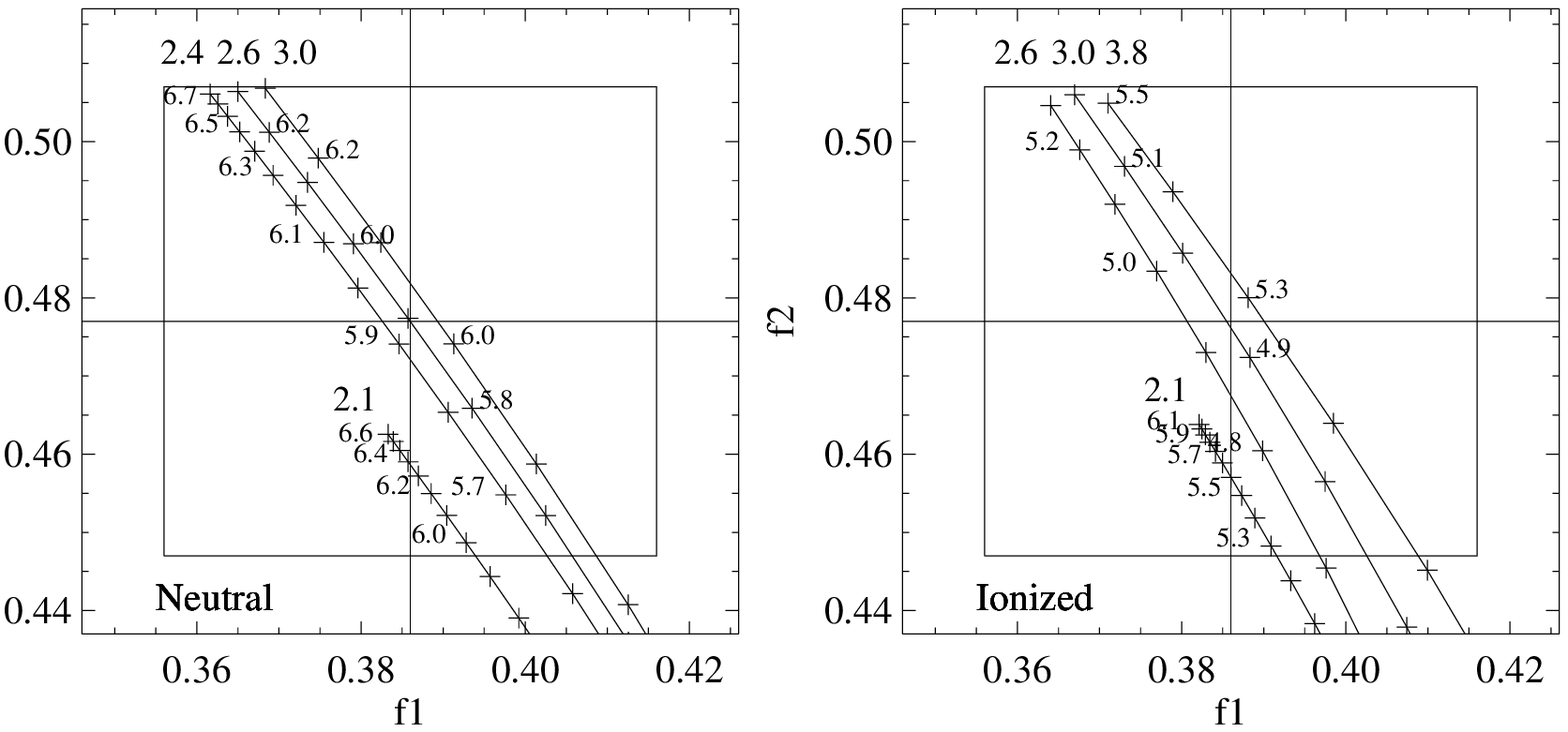}}
\figcaption[f13.eps]{\label{f2f1}
The ($f1,f2$) planes for the expected ratios of the column densities of
the  \ion{C}{1} fine structure states, assuming (left panel) neutral
hydrogen impactors and (right panel) ionized impactors.  The curves are
the loci of ($f1,f2$) ratios for different pressures at a constant
temperature.  The large numbers to the upper left of each curve are the
logarithms of the temperature.  The log of the pressure in units of
$\log(n_{h}T) =\log(P/k)$  is printed for every other density point
along the loci of points for each constant temperature curve.  The region
shown is centered on ($f1,f2$) = (0.386, 0.477).  The box centered on
this point is the error limit  ($\sigma_{f1},\sigma_{f2}$) = (0.04,
0.04). Despite the large uncertainty there is nearly an order of magnitude
difference in the pressure for the two limiting cases.  The temperature
for the ionized impactors is closer to that of
\Htwo.
 }
\end{figure*}

\subsubsection{CO Upper Limits}

\citet{Bachiller:2000} presented a detailed contour map of the CO
(2--1) emission in M27.  The contours fall steeply towards the CS
where, using the conversion factor suggested by \citet{Huggins:1996},
we estimate an upper limit for the CO column of 10$^{15}$ cm$^{-2}$
along the line-of-sight.  CO has a number of strong absorption bands in
the far UV, such as the numerous A-X bands spread throughout the STIS
E140M bandpass, and this column density would easily  be detected.  In
Figure~\ref{co} we show the E140M spectral region surrounding the A-X
(2-0) 1477.565 \AA\ bandhead.  In the top panel we overplot CO
absorption models for representative excitation temperatures of 3000 K
and 300 K at a column density of 10$^{15}$ cm$^{-2}$ in blue and red
respectively.  We  assumed the doppler parameter is the same as for
\Htwo, 6.5 \kms. Clearly the column density is much lower
than 10$^{15}$ cm$^{-2}$.  In the lower panel we show  models for 3000
K with a column density of 4 $\times$ 10$^{14}$ cm$^{-2}$ and for  300
K with a column density of 8 $\times$ 10$^{13}$ cm$^{-2}$.  The
structure in the models is on the order of the noise in the
observation, so we will use these values as the (excitation temperature
dependent) upper limits for line-of-sight CO.  Using a lower excitation
temperature would yield an even lower the upper limit.

\subsubsection{\ion{C}{1} Fine Structure Pressure Diagnostic} 
\label{cifine}

Although CO was not detected we do detect excited \ion{C}{1}, \ion{C}{1}* and \ion{C}{1}** multiplets in the E140M data with a velocity of -75~\kms, coincident with the velocity of the excited \Htwo.
The level populations of the three fine structure levels in the ground
of \ion{C}{1}, assuming a long enough time has passed for equilibrium
to be established, is a detailed balance of the de-excitation and
excitation rates set by collisions of \ion{C}{1} with other particles
(the impactors -- \ion{H}{1}, \ion{He}{1}, p, e$^-$, ortho-H$_2$,
para-H$_2$) and the radiative decay rates of the fine structure levels.  \citet{Jenkins:1979}
present a useful diagnositic for gas pressure based on the ratios of
the column densities for $N_{CI*}$ and $N_{CI**}$ relative to the total
column density $N_{tot}$= $N_{CI}$ + $N_{CI*}$ + $N_{CI**}$, (f1 =
$N_{CI*}$/$N_{tot}$ and f2 = $N_{CI**}$/$N_{tot}$).  The pressure is
constrained by comparing the measured values of f1 and f2 against
theoretical loci for $f1$ and $f2$ at variable temperature and 
density.

We determined $f1$ and $f2$ by $\chi^2$ fits to the absorption profiles
of the \ion{C}{1} \dlam1656.27 -- 1658.12  UV2 multiplet  using the
\citet{Morton:2003} oscillator strengths. The result is shown in Figure
\ref{cifit}.   We determined a doppler parameter of $b$ = 4.5 \kms\ and
we find $N_{CI}$ = 3.0 $\pm$0.5 $\times$ 10$^{13}$ cm$^{-2}$, $N_{CI*}$
= 8.5 $\pm$0.5 $\times$ 10$^{13}$ cm$^{-2}$ and  $N_{CI**}$ = 10.5
$\pm$0.5 $\times$ 10$^{13}$ cm$^{-2}$, yielding ($f1,f2$) = (0.386,
0.477) $\pm$ 0.04.

The theoretical loci for $f1$ and $f2$ depend on the mix of impactors
assumed.  \citet{Jenkins:2001} discuss three extremes: Case1 where all
the impactors are neutral hydrogen, Case 2 where all the hydrogen is
molecular and Case 3 where all the hydrogen is ionized. They find Case
2 and Case 1 to yield very similar loci in the $(f1,f2)$ plane.  Using
their code (kindly provided by Jenkins, private communication) we show
the results for Case 1 (left) and Case 3 (right) in Figure~\ref{f2f1}. The crosshairs mark the  ($f1,f2$) point and the figure ranges over the 
$\pm$ 0.04 error limits.  Crosses
mark the density at intervals of 0.1 dex.  The log of the pressure, in
units of $\log(n_{h}T) =\log(P/k)$,  is printed for every density point
along the loci of points for each constant temperature.  

For the neutral case the (0.386, 0.477) point lies on the $\log(T)$ =
2.6 curve with a pressure of $ \log{n_{h}T} $ = 5.9, while
the ionized case has $\log(T) \approx$ 3.0 with a pressure of
$4.9 < \log{n_{h}T} < 5.0 $.  The allowed range of pressure for the
$\pm$ 0.04 error box in the neutral case is 5.7 $
\lesssim\ \log{n_{h}T} < \infty $ and for the ionized case 4.8 $
\lesssim\  \log{n_{h}T} < \infty $ with the $\infty$ density case being
associated with the coldest allowable logarithmic temperature of
$\approx$ 2.1 dex.\footnote{For constant temperture curve the ($f1,f2$)
points converge to a point as the logarithmic density approaches 6 and
the diagnostic becomes less discriminatory.} 
\begin{figure}
\centerline{\includegraphics[height=8.5in]{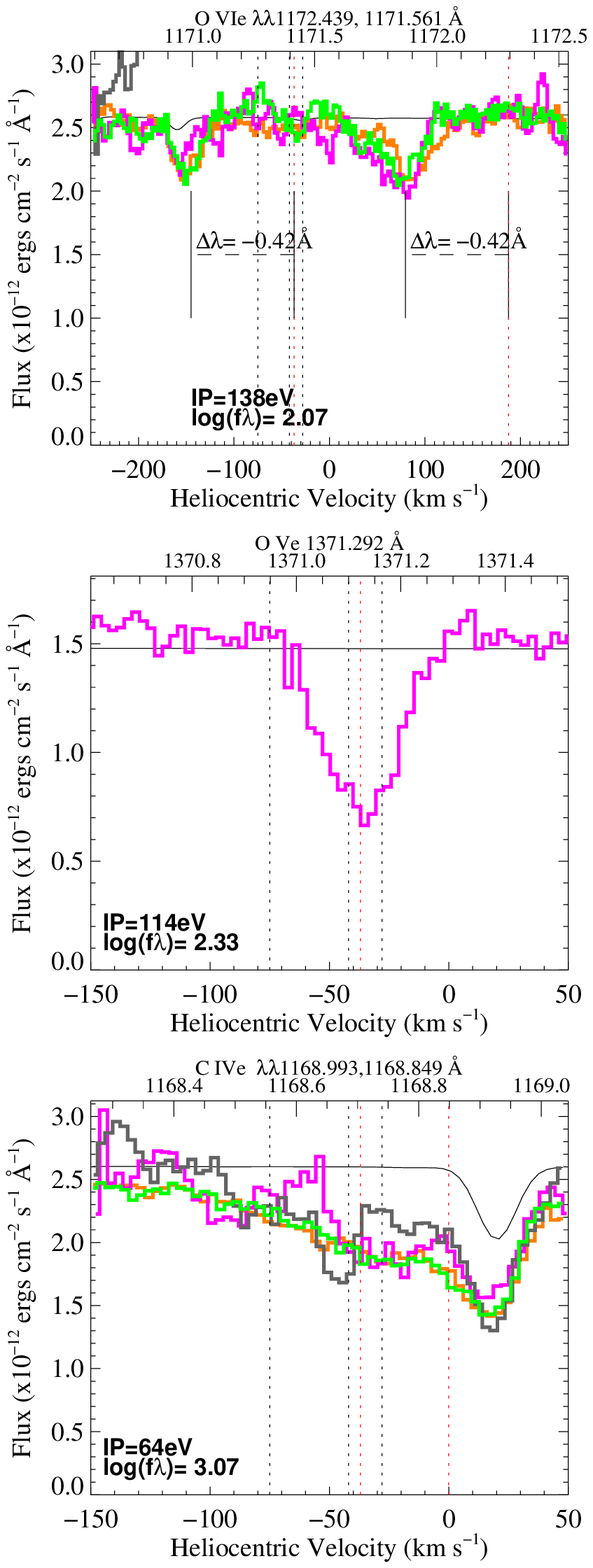} \vspace*{-.15in}}
\figcaption[f14.eps]{ \label{oviovcive}
\ion{O}{6}*, \ion{O}{5}* and \ion{C}{4}* lines used to reconcile
the \FUSE\ and STIS wavelength offsets.  \ion{O}{5}e
\lam1371.292  establishes the gravitational redshift zero point.
\ion{O}{6}e and \ion{C}{4}e are observed by both \FUSE\ and STIS.  \FUSE\ spectra s12 (orange), s21 (green) and the model (black) are plotted along with 
overlapping STIS orders (purple and grey), if available.
Vertical dashed lines mark the transition, $V_{sys}$, $V_{gr}$ and
non-nebular velocities at --75, --42, --37 and --28
\kms\ respectively.  Red lines near 0 \kms\ and +190 \kms\ in the respective \ion{C}{4}e and \ion{O}{6}e panels mark
gravitational redshift of redward doublet member. \ion{O}{6}e wavelength discrepancy (--0.42 \AA) with NIST is noted.
}
\end{figure}

The coincidence of the
\ion{C}{1} velocity with the \Htwo\ velocity argues that
these two species are co-located.  The $\sim$ 2000 K implied by the
\Htwo\  ro-vibrational distribution is closer to the temperature given by
the ionized case than the neutral case.  This suggests the excited
\ion{C}{1} and \Htwo\ are located in a warm and electron
rich medium with $n_e \sim $ 80 cm$^{-3}$.  In this 
environment it may be possible for \Htwo\ to form at the
interface between the ionized medium and the PDR clumps by   H + e$^-$
$\rightarrow$ H$^-$ + $h\nu$ followed by H + H$^-$ $\rightarrow$ H$_2$
+ e$^-$ mechanism, \citep[c.f.][]{Aleman:2004, Natta:1998}.  We note
that our total column density for the neutral carbon,
$N_{CItot}$ = 2.2  $\times$ 10$^{14}$ cm$^{-2}$ is $\sim$ 2 orders of magnitude
higher than what is expected for $N_{CItot}$ in typical \ion{H}{2} regions,
as discussed in Appendix A of \citet{Jenkins:1979}.  It is likely
that the source of \ion{C}{1} in the diffuse medium is CO dissociation in the clumpy medium.

In addition to UV2, we also examined the fits for the UV3, UV4 and UV5
multiplets.  We found that the $\chi^2$ fits to these multiplets
yielded slightly different results.  In the cases of UV3 and UV5, there were intervening unidentified absorption features
that caused the $\chi^2$ to drift to higher columns for the $N_{CI**}$
lines. In the case of UV4 the column densities were different but the $f1$
and $f2$ ratios were essentially the same.  Currently there are disagreements in
the literature \citep[c.f.][]{Morton:2003,Jenkins:2001} regarding the
relative strengths of the higher multiplets with respect to UV2.  Since
the resulting $f1 and f2$  ratios all lie within the  $\pm$ 0.04 error
box, we decided to rely upon the multiplet with the most certain
oscillator strengths, UV2.  Our conclusions that the \ion{C}{1} is excited, the column density is high compared to typical \ion{H}{2} regions,  and  the temperature implied by the ionized impactors are closer to that implied by the
\Htwo\ level population, are unchanged by our reliance on UV2 alone.

\subsection{ \FUSE\ and  STIS Absorption Profiles -- Ionization Stratification}
\label{profiles}

The metal line profiles show absorption  from velocity components that
we associate  with the stellar photosphere located near the systemic
velocity of the star, the blueshifted nebular outflow, and the
redshifted non-nebular ISM.  In general, the high ionization species
tend to appear with photospheric and nebular velocity components while
the low ionization and neutral species tend to show nebular and
non-nebular components.  A gravitational redshift of $\approx$ 5
\kms\ can be seen in the high excitation stellar \ion{O}{5}e \lam 1371
line. (We use the ``e'' specification with the atomic species designator to
indicate transitions to lower levels other than the ground
state).

\begin{figure*}[t]
\centerline{\includegraphics[height=8.5in]{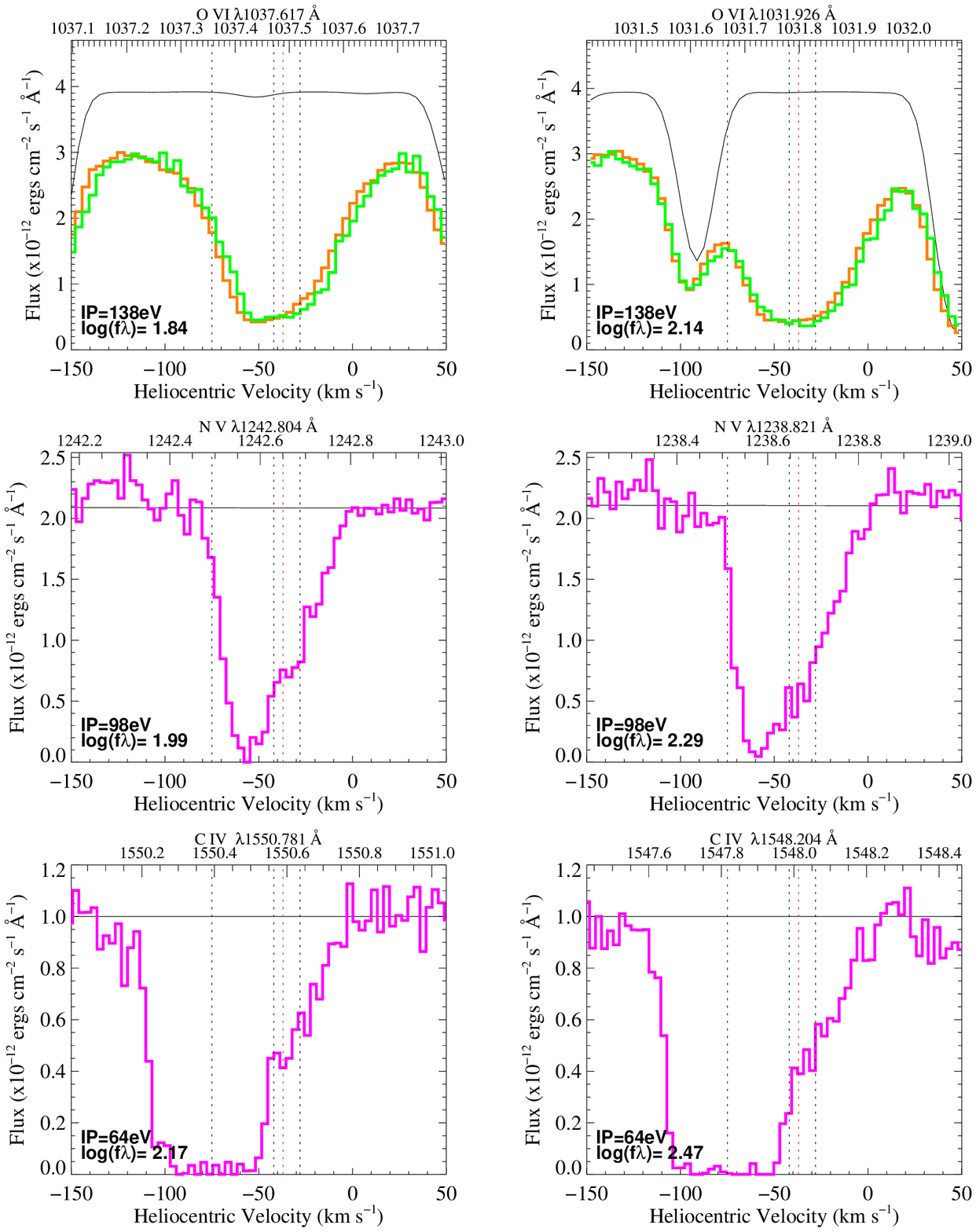}}
\figcaption[f15.eps]{ \label{ovinvciv}
Absorption as a function of velocity for the high ionization resonance lines of \ion{O}{6}, \ion{N}{5}, and \ion{C}{4}, showing varying degrees of photospheric broadening to the red.  Nebular outflow to the blue of systemic increases with decreasing ionization potential. 
See Figure~\ref{oviovcive} for a description of the colors. }
\end{figure*}

\begin{figure*}
\centerline{\includegraphics[height=8.5in]{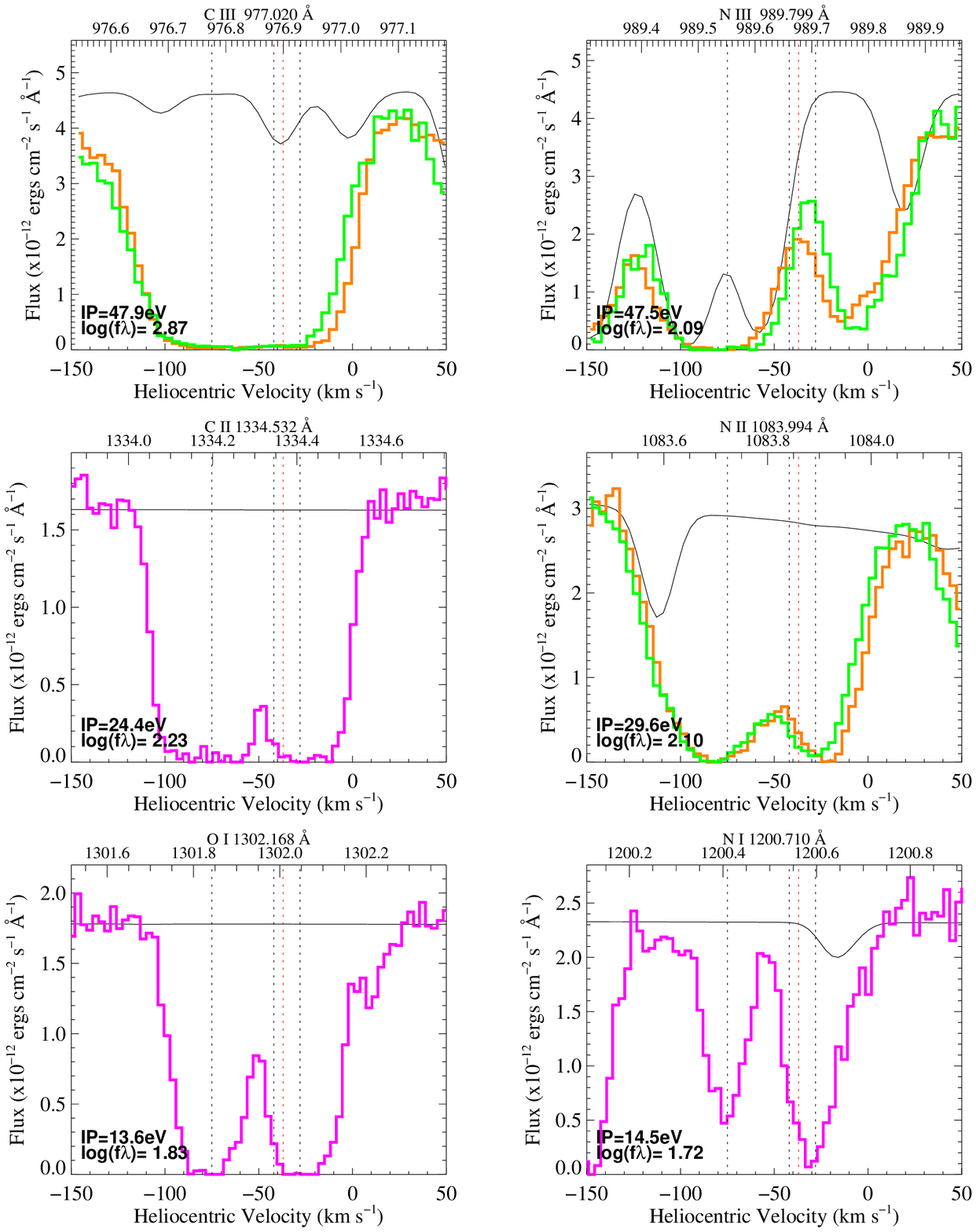}}
\figcaption[f16.eps]{ \label{cno}
Absorption as a function of velocity for representative \ion{C}{3}, \ion{C}{2} 
\ion{O}{1}, \ion{N}{3}, \ion{N}{2} and \ion{N}{1} lines.
Ions are fully saturated throughout the
nebular flow region blueward of  --42 \kms.  Neutrals are less so and favor
strong absorption near --75 \kms.  See Figure~\ref{oviovcive} for
a description of the colors.   }
\end{figure*}

\begin{figure*}
\centerline{\includegraphics[height=8.5in]{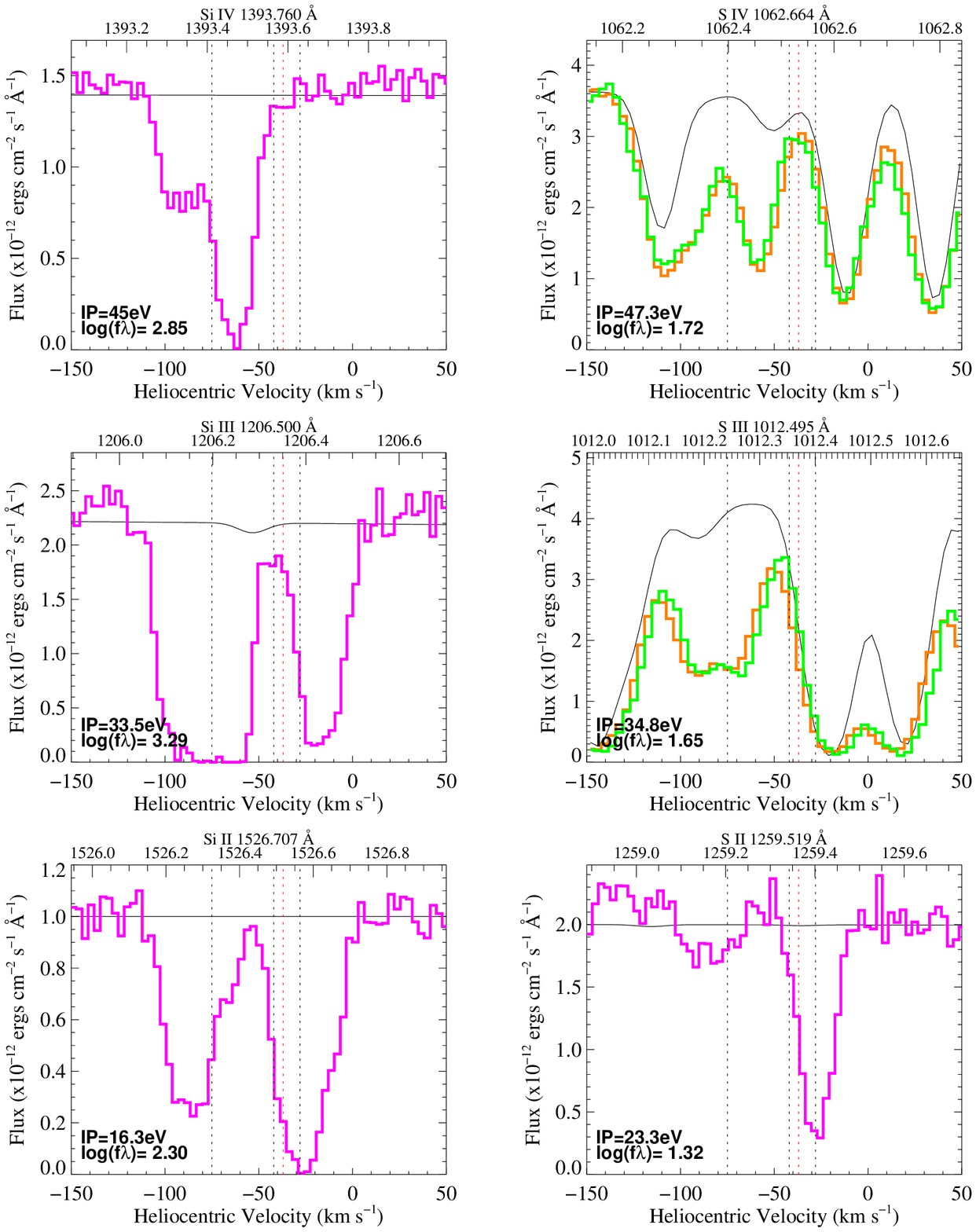}}
\figcaption[f17.eps]{\label{sis}
Comparision of the \ion{Si}{4} -- 2 to \ion{S}{4} -- 2.  The low
ionization species, \ion{Si}{2} and \ion{S}{2}, dominate in the zone
between --75 and --110 \kms\ while the high ionization species,
\ion{Si}{4} and \ion{S}{4}, dominate  near --60 \kms. Transition in
ionization occurs where absorption by the neutrals and \Htwo\ is
strongest.  See Figure~\ref{oviovcive} for a description of the
colors.
 } 
\end{figure*}

We show a select subset of the metal line absorption
profiles in  Figures \ref{oviovcive} -- \ref{sis}.  \FUSE\ ~s12 and s21
spectra are shown in orange and green respectively, while the STIS
lines are plotted in purple (and grey if two orders contain the same wavelength range). The continuum model, including the
hydrogen lines, is plotted as a thin black line. The heliocentric reference frame is used. The black dashed
vertical lines mark the location of the hot \Htwo\
component at --75 \kms\, the systemic velocity at --42 \kms, and the cold
\Htwo\ component at --28 \kms.  The red dashed vertical line
marks the gravitational redshift of the systemic velocity.  To convert from the nebular rest frame, subtract the heliocentric velocity from the systemic velocity.

\subsubsection{Stellar Photospheric Features and the Absolute
Wavelength Scale}
\label{abslam}

Comparison of the \FUSE\ and STIS wavelength scales revealed a
systematic offset.  The reconciliation of this offset is essential for
investigating the kinematics of the nebular outflow, where we seek to
determine the velocity of the various molecular and atomic features
with respect to the systemic velocity of the nebula. 
Lines that arise from the photosphere should match the systemic
velocity of the system less the offset caused by gravitational redshift.  
An absolute reference to the heliocentric velocity was established by close examination of the \ion{O}{5}e \lam 1371.296 feature, shown in the middle panel of Figure~\ref{oviovcive}.
This narrow line results from a transition between two highly excited states in
\ion{O}{5}e and is expected to be an excellent indicator of the
photospheric restframe (Pierre Chayer private communication).  For a
compact object of this mass and radius (\S~\ref{previous}) we expect
the photospheric lines to experience a gravitational redshift,
$V_{gr}=c((1-\frac{2GM_{*}}{R_{*}c^2})^{-\onehalf}-1$) = 5.1~\kms.
Applying a shift of --13 \kms\ to the STIS spectrum placed the centroid
of the \ion{O}{5} \lam 1371.296  at --37 \kms, as  expected for a
$V_{sys}$ of --42 \kms\ \citep{Wilson:1953}.

Examination of overlapping \FUSE\ and STIS spectra in the wavelength
regions below 1190 \AA\ revealed some systematic disagreements.  In the
original analysis of the \FUSE\ M27 spectra by \citet{McCandliss:2001a}
the hot nebular \Htwo\ component was defined to be at --69 \kms.  We
found it necessary to shift the \FUSE\ spectra blueward by --6 \kms,
such that the hot nebular \Htwo\ is now at --75 \kms.  The most useful
overlap lines for assessing the alignment were the doublet blend of
\ion{C}{4}e \dlam 1168.849, 1168.993, bottom of Figure~\ref{oviovcive},
and the the narrow \ion{O}{6}e \dlam1171.56, 1172.44, top of
Figure~\ref{oviovcive}. We note the wavelengths of \ion{O}{6}e doublet
given in the National Institute of Standards and Technology (NIST)
online tables (\url{http://physics.nist.gov/PhysRefData/ASD/lines\_form.html})
appear to be in error by $\approx$  --0.42 \AA\ \citep[see][]{Jahn:2006}.   These spectra are comparatively noisy,
but the alignment with the \FUSE\ spectra agrees as well as the
alignment between the s12 and s21 spectra.  We also used  the
\FUSE\ \ion{N}{1} multiplets at \dlam 1134 -- 1135 and the STIS
\ion{N}{1} multiplets at \dlam 1200 -- 1201 along with excited
\Htwo\ lines that appear in the STIS bandpass above 1190 \AA\ with the
continuum plus hydrogen model of \citet{McCandliss:2007} to assess the
consistency of the wavelength reconciliation. The error in the systemic velocity (\S~\ref{previous}) is of order the
\FUSE\ resoution element and is three times the STIS resolution
element.  We consider the agreement of line profiles from spectra
acquired with two different instruments to be excellent.

\subsubsection{Photospheric + Nebular Features}

\paragraph{\ion{O}{6}, \ion{N}{5} and \ion{C}{4}}

Figure~\ref{ovinvciv} shows the high ionization resonance doublets of
\ion{O}{6} \dlam1032.62, 1037.62, \ion{N}{5} \dlam 1238.821, 1242.804
and \ion{C}{4} \dlam 1548.204, 1550.781.  All these lines show
photospheric absorption to the red of the systemic velocity.  The
strength and terminal velocity of the blue shifted portion of the line
profiles increases with decreasing ionization potential.  The
\ion{O}{6} lines show only slight signs of blue shifted nebular
absorption.  The nebular absorption component in the \ion{N}{5} lines
is strong only between --42 and --75 \kms\ just reaching saturation at
$\approx$ --60 \kms.  In contrast, the nebular absorption component in
the \ion{C}{4} lines spans --42  to --115 \kms\ and is completely
saturated from --50  to --95 \kms.

\subsubsection{Intermediate Ionic and Neutral Nebular Absorption}

\paragraph{\ion{C}{3} -- \ion{}{2}, \ion{N}{3} -- \ion{}{1} and \ion{O}{1}} 
Figure~\ref{cno}
show lines of \ion{C}{3} -- \ion{}{2}, \ion{N}{3} -- \ion{}{1} and \ion{O}{1}.  Like the \ion{C}{4} lines, the \ion{C}{3} -- \ion{}{2} lines are heavily saturated throughout the
nebular flow region blueward of  --42 \kms.   The saturation
makes it difficult to tell whether any one ion is dominant in the nebular outflow.
\ion{N}{3} is strongly blended  with overlapping \Htwo\ features. However,
it shows saturated absorption between -60 and --100 \kms. 
\ion{N}{2} \lam 1083.994 shows absorption throughout the flow, being less saturated at low velocities and becoming completely  saturated at --75 \kms.
This line also shows stronger nebular absorption than non-nebular absorption.
In contrast, \ion{N}{1} \lam 1200.223 shows weaker nebular absorption than non-nebular absorption.  Both \ion{N}{1} \lam 1200.223 and \ion{O}{1}
show an absorption component centered on --75 \kms, which coincides with the velocity of the absorption maximum in \Htwo\ and \ion{C}{1}.

\paragraph{\ion{Si}{4} -- \ion{}{2}, and \ion{S}{4} -- \ion{}{2}}
The transition zone between high ionization and low ionization occurs
at the velocity of --75 \kms\ where \ion{H}{1}, \ion{C}{1}, \ion{N}{1},
\ion{O}{1} and \Htwo\ show up most strongly in the nebula.
The high ionization -- low velocity, low ionization -- high velocity
dichotomy is best illustrated by examining the velocity profiles of the
low abundance metals Si and S in Figure~\ref{sis}.  \ion{Si}{2} is
stronger than \ion{Si}{4}  in the zone between --75 and --110 \kms, while
\ion{Si}{4} is stronger than \ion{Si}{2}  in the zone between --42 and
--75 \kms.  The \ion{Si}{3} \lam 1206.500 line is saturated throughout
most of the flow.  The \ion{S}{4} -- \ion{S}{2} ions show similar behavior.

\section{Discussion}
\label{discuss}

Here we enumerate the main findings from the analysis.

\begin{enumerate}

\item The nebular \Htwo\ is highly excited with a ground state ro-vibration population that
deviates significantly from the best fit single temperature Boltzman
distribution of 2040 K. The total column density is 7.9
$\times$ 10$^{16}$ cm$^{-2}$.  The deviations are characterized by
a flatter slope for the lower rotational levels (0 $\le\ J^{\prime\prime} \lesssim $ 5)
than for the higher rotational levels (6 $\lesssim\ J^{\prime\prime} \le $ 11).

\item Continuum fluorescence of \Htwo\ has not
been detected to the background limit of the sounding rocket
observation $\approx$ 5 $\times$ 10$^{-17}$~\bright.  However,
Ly$\alpha$ fluorescence  has been
detected by \citet{Lupu:2006}.  This emission mechanism is possible only 
for \Htwo\ with a significant population in the \vpp\ = 2 vibrational state in  the presence of a strong \lya\ radiation field.

\item  The stellar SED exhibits negligible stellar reddening along the
line-of-sight.  This finding is at odds with the variable reddening
indicated by the ratio of \ha/\hb\ intensity as reported in the
literature and confirmed in this study in the immediate vicinity of the
star.  We note that the reddening determined by \hg/\hb\ and
\hd/\hb\ is consistent with zero and conclude it is possible that
another physical process, aside from reddening by dust, is causing the
discrepant \ha/\hb.

\item The nebular \Htwo\ appears at a heliocentric velocity of --75
\kms,  corresponding to an expansion velocity of 33
\kms\ in the nebular  restframe.  \ion{H}{1}, \ion{C}{1}, \ion{N}{1},
and \ion{O}{1} exhibit strong resonance line absorption at the same
expansion velocity.  The molecules and neutrals demarcate a transition
velocity between a regime of high ionization and low expansion velocity
(10 $\lesssim\ v_{hi} \lesssim $ 33 \kms) and a regime of low ionization
and high velocity (33 $\lesssim\ v_{low} \lesssim $ 65 \kms).  The upper
limit to the terminal expansion velocity is $\approx$ 70 \kms.  These
observations, along with the absence of any signs for a high velocity
radiation driven wind, are a challenge to interpret in the context of the simple interacting winds model for PN formation \citep{Kwok:1978}.
\citet{Meaburn:2005} \citep[see also][]{Meaburn:2005a,Meaburn:2005b}, who used high resolution position-and-velocity spectroscopy of the optical emission
lines to derive the ionization kinematics of several objects, have
arrived at a similar conclusion.  They find that ballistic ejection
could have been more important than interacting winds in shaping the
dynamics of PNe.  \citet{Balick:2002} have reviewed in detail
some of the more sophisticated thinking about the dynamical processes that shape PNe.

\item  Analysis of the \ion{C}{1} 
multiplet fine structure populations gives better agreement with the
expected temperature of $\sim$ 2000 K for collisions with protons and electrons as opposed to collisions with neutrals or molecules.  This suggests the
molecular and neutral transition component is in close contact with ionized material.

\item The upper limits to CO in the diffuse line-of-sight gas range over  8 $\times$ 10$^{13}$ cm$^{-2}$ $< N(CO) <$ 4 $\times$ 10$^{14}$ cm$^{-2}$ depending on the assumed CO rotational excitation temperature.  The high column density of
\ion{C}{1} found in the nebula is consistent with it being a daughter product
of CO dissociation.

\item The lower limit to the neutral to \Htwo\ column density ratio is $N$(\ion{H}{1})/$N(H_2)$ = 127 in the transition region (using component number 3 in Table~\ref{HIcolbvel}).

\end{enumerate}

We will now discuss the constraints imposed by  these findings on the  ionization kinematics, mass structures, nebular dust,  and the excitation, formation and destruction
of \Htwo\ within the nebula.

\subsection{Ionization Kinematics} 
\label{ksri}

Low mass to intermediate mass stars begin to evolve into AGB stars when
core nucleosynthesis ceases and a degenerate core forms.  The AGB phase
lasts $\approx$ 250,000 years and is marked by a series of high mass
loss events that merge to produce a slow wind.  At some point the
star collapses into a hot compact degenerate object.  In the
PN formation scenario suggested by \citet{Kwok:1978},  a fast
($\sim$ 1000 \kms) low density radiation driven wind from the hot star
shocks and ionizes the slow moving ($\sim$ 10 \kms), high density
mostly molecular AGB wind, resulting in the expanding structures we see at 
latter times.

\citet{Natta:1998} have computed detailed time dependent changes
in the atomic and molecular emissions in a  phenomenologically
motivated interacting wind model.  Their model consists of an interior
ionization region bounded by a shell propagating outward with a
velocity of 25 \kms\ overrunning and shocking a mostly molecular AGB
wind with a velocity of 8 \kms.  An ionization-bounded time-dependent
PDR results with three layers.  These are,  a fast moving ionized shock interface,
an  adjacent dissociation layer, and a slow moving  molecular layer
further downstream.  They find most of the shell mass to be
vibrationally excited \Htwo\ and \ion{H}{1} production to
be modest.  After $\approx$ 4000 years  they find the infrared
\Htwo\ line diagnostic of  S(1) (1--0) / S(1) (2--1)
indicates a switch from thermal excitation to fluorescent excitation.
They also find that \Htwo\ does not co-exist with CO, as
dissociation favors C$^{+}$ + O.  Consequently, they suggest molecular masses
calculated for old PN based on CO emission and standard CO/H$_2$ ratios
may be underestimated.

M27 has a kinematic age of $\approx$ 10,000 yrs \citep{O'Dell:2003}, so
perhaps it is reasonable that the nebula we see today bears little
resemblance to this simple layered, two-velocity interacting wind
picture.  We see no evidence for a shell of predominately molecular
material along the diffuse line-of-sight medium, nor do we see neutral
or molecular material at low expansion velocity as would be expected
for an unperturbed AGB wind.  The upper limit to the molecular
fraction\footnote{By nucleon number $f_{H_2}$ $\equiv$
2$N(H_2)$/(2$N(H_2)+N(H)$) with $N(H)$=$N$(\ion{H}{1})+$N$(\ion{H}{2}).} in
the neutral velocity component expanding at 33 \kms is $f_{H_2}$ =
0.016, assuming $N$(H)=$N$(\ion{H}{1}).  We do see evidence for velocity
structure in the intermediate ionization states (Figures~\ref{cno} --
\ref{sis}) and \ion{H}{1} (Table~\ref{HIcolbvel}) out to expansion
velocities of $\approx$ 65 \kms, but there is no evidence for P-Cygni
profiles with $\sim$ 1000 \kms\ terminal velocities in any of the high
ionization lines (Figure~\ref{ovinvciv}).
 
We find instead a rather permuted velocity structure of neutral and
molecular material embedded (in velocity space) between the high and
low ionization medium.  Our findings suggest that the AGB wind has been
accelerated through interaction with the hot star and that much of the
original mostly molecular atmosphere  has been almost
completely dissociated.  The surviving molecular material is currently  interacting strongly with the highly ionized
medium and we postulate that the intermediate ionization species are
created and accelerated in this interaction.

\citet{Villaver:2002b} have treated the radiative hydrodynamical
problem with great rigor for a range of progenitor masses, using as a
starting point for their PN simulation the structure found at end of
AGB evolution by \citet{Villaver:2002a}.  They find a main nebular
shell develops surrounded by a relatively unperturbed halo. They give
detailed figures for the evolution of density, velocity and
\ha\ emission brightness.   We note that the amplitude of the velocity
structures for their 1 -- 2.5 M$_{sun}$ models are in the range of the
outflow velocities exhibited by the line profiles in
Figures~\ref{ovinvciv} -- \ref{sis}.  Unfortunately they give no
details on the ionization kinematics of the main and halo regions as a
function of time, as \citet{Natta:1998} have done.  A model combining the
detailed time dependent calculation of the atomic and molecular emissions of \citet{Natta:1998} with the radiative hydrodynamical rigor of \citet{Villaver:2002b} would be useful for interpreting these observations.

\subsection{ Mass Structures }

Identifying kinematic components appearing in spectra with spatial
structures in images must be approached with care.  The computed total
mass of these velocity components will depend on the assumed diameter
and thickness of  particular structures in the nebula.  However, it
seems reasonable to expect the 33 \kms\ transition zone, where most of
the molecular and much of the neutral hydrogen appears, to be at or near the
region where the \ha\ brightness drops to near zero.  Inside this zone
we expect the medium to be completely ionized.  The minor axis of
the bright \ha\ in M27 is $\approx$ 300$^{\prime\prime}$ and the full
diameter including the faint halo detected by
\citet{Papamastorakis:1993} is 1025\arcsec.  Assuming a distance of 466
pc gives radii for the transition region and the total nebular extent
to be $\approx$ 0.34 and 1.16 pc respectively.  These radii agree
reasonably well with the bright \ha\ emitting region and the full extent of
the AGB wind found in the \citet{Villaver:2002a,Villaver:2002b}
simulations after $\approx$ 10,000 years .

We estimate the masses of the various structures given the following
(gross) assumptions.  Let the ionized region be a sphere with a radius
of 0.34 pc and an electron density of $\sim$ 300 $\pm$ 200 cm$^{-3}$
\citep{Barker:1984}.  The ionized mass becomes  0.4 $\lesssim M_{ion}
\lesssim$ 1.9 M$_{\sun}$.  If  the total neutral hydrogen column
density (3 $\times$ 10$^{19}$ cm$^{-2}$) is uniformly distributed in a
shell between 0.34 and 1.16 pc with a mean density of 12 cm$^{-3}$,
then $M_{sh}$ = 1.7 M$_{\sun}$.  Our constraint on the progenitor mass,
the sum of the ionized, shell and degenerate star masses, is 2.6
$\lesssim M_{pro} \lesssim$ 4.2 M$_{\sun}$.

By comparison, the molecular material in the diffuse nebular medium has a much
lower mass.  Although the \Htwo\ is associated with only
one neutral hydrogen velocity component, we assume it is spread
uniformly throughout the 0.82 pc thick shell.  Using this maximal
volume yields an upper limit on the molecular mass in the diffuse
medium, $M_{dmol} \lesssim $ (1.7 M$_{\sun}$)/375  = 4 $\times$
10$^{-3}$ M$_{\sun}$.  \citet{Meaburn:1993} have estimated the total
mass of the eleven largest globules, which appear as dark knots against
the bright nebular \ion{O}{3} emission, to be $M_{omol}$ = 1.1 $\times$
10$^{-3}$ M$_{\sun}$.  They used  measurements of \ion{O}{3}
transmission and assumed the general ISM gas-to-dust ratio
 \citep{Bohlin:1978}.  The Meaburn et al. clump mass
estimates compare reasonably well to those of \citet{Huggins:1996} who
estimate the total molecular mass associated with the CO emitting
clumps to be $M_{cmol}$ = 2.5 $\times$ 10$^{-3}$ M$_{\sun}$.  They
assumed a thin ring geometry  and a fraction of CO/H$_2$ = 3 $\times$
10$^{-4}$, a value $\ge$ 10 times than the highest found by \citet{Burgh:2007}
for the diffuse ISM.  Use of a lower ratio will increase the CO clump mass estimate as noted by \citet{Natta:1998}.

Our upper limit to the diffuse molecular mass is on the same order as the clump mass estimates.   The upper limit on the total
molecular mass in the nebula, the sum of the diffuse and CO
derived masses (assuming the high CO/H$_2$ ratio),  is  M$_{mol}$ = 6.5 $\times$ 10$^{-3}$ M$_{\sun}$.  We conclude that there is no large mass of molecular material
in M27, either in the diffuse or clumped medium.  This is in contrast to
the  \citet{Meixner:2005} finding for the Helix, for which they estimate the mass of all the \Htwo\ emitting globules to be
$\sim$ 0.35 M$_{\sun}$.  We note that all of the molecular mass estimates
for M27 are on the order of the mass of the planets in our own solar
system, $\approx$ 1.3 $\times$ 10$^{-3}$ M$_{\sun}$.  The suggestion
that dense circumstellar bodies have survived the AGB phase and
may constitute a significant fraction of the molecular material in the
nebula environment cannot be ruled out.
 
The absence of CO absorption in the diffuse medium is consistent with
the \citet{Natta:1998} finding that CO and \Htwo\ are
unlikely to co-exist in the gas phase, because CO is quickly
dissociated after leaving the vicinity of the clumps.
\citet{Huggins:2002} confirmed this picture with exquisite observations
of the globules of the Helix.  They show how CO survives only in the
shadows of the clumps. The heads of these globules show  strong
coincident emission of H$_2$ S(1) (1--0) and \ha, and weaker coincident
emission in the shadowed zone.  \citet{O'Dell:2003} note the globules
in M27 tend towards a more irregular and tail-less appearence than
those in the Helix.  Nevertheless the observation of H$_2$
emission in the vicinity of the globules suggests they may be a source of H$_2$ for the diffuse nebular medium.

The temperature of the diffuse medium inferred from the molecular
hydgrogen ro-vibration levels exceeds the sublimation temperature of
refractory dust, $\sim$ 1500 K \citep{Glassgold:1996}.  While it is
unlikely that the gas and dust will be in thermodynamic equilibrium,
the absence of any significant reddening of the central star SED
suggests that the diffuse environment is as inhospitable to dust as it
is to CO and \Htwo.  Dust, or perhaps even compact bodies, may exist in
the CO rich dense clumps, which we will argue in \S~\ref{clumpsource},
are the likely source of \Htwo\ for the diffuse medium.

\subsection{ Constraints on the Excitation, Formation and Destruction of \Htwo }
\label{fordest}

\subsubsection{\Htwo\ Excitation Processes}
\label{h2excite} 

The models of \citet{Natta:1998} suggest that in the latter stages of
planetary nebula evolution the \Htwo\ is being excited via far-UV
continuum fluorescence, yet we have failed to detect this emission.  We
briefly review this process.

\Htwo\ is excited from the ground vibrational level of
\xsig\ into higher electronic states, predominately  \bsig\  and
\cpi\ (the Lyman and  Werner bands), by the absorption of a far-UV
photon (912~$\le$~$\lambda$~$\la$~1120 \AA) emitted by a star.  The
short wavelength cutoff is imposed by the usual assumption that enough
neutral hydrogen is present to shield \Htwo\ from
excitation by stellar Lyman continuum (\lyc)
photons.   The rate of molecular excitation is proportional to the
stellar flux absorbed by the ro-vibrational population of the ground
electronic state \xsig.  Fluorescence follows excitation, producing a
highly structured emission in the 912 -- 1650 \AA\ bandpass,
as electrons fall back into the excited vibrational levels  of \xsig.
Approximately 11 - 15$\%$ of the time, the fluorescent pumping process
leaves the molecule in the  vibrational continuum of the ground state (\xsig),
located 4.48 eV above the (\Jpp,\vpp) = (0,0) level,\footnote{Direct radiative dissociation of the \xsig\ state is dipole forbidden. } resulting in
its spontaneous dissociation into $H(1s) + H(1s)$ \citep{Stecher:1967, Dalgarno:1970, 
Draine:1996}.  Electrons ending up below the vibrational continuum
cascade toward the ground vibrational state \vpp\ = 0
via slow quadrupole transitions with radiative lifetimes $\sim$ few days and longer \citep{Wolniewicz:1998}, producing the infrared fluorescence
spectrum  \citep{Black:1976}.  

Lyman and Werner band absorption of the central star continuum
unavoidably results in ultraviolet fluorescence, so why was it not
detected?  To answer this question we estimate the total brightness by simply calculating
how many stellar photons are absorbed by \Htwo\ at a given
distance, which we take to be the transition zone, and
re-emitted into 4$\pi$ steradians.  This will give an upper limit to
the total brightness of the \Htwo\ in the diffuse medium.
We will neglect absorption by dust.

Using our hydrogen absorption model we find 30\% of the photon flux in the
912~--~1120~\AA\ bandpass is absorbed.  Reradiation into 4$\pi$
steradians at 0.34 pc  yields a total brightness of 3.4
$\times$~10$^{-5}$~\brightnoaast (assuming an average energy per photon
of 1.6  $\times$ 10$^{-11}$ ergs).  Neglecting the fluorescence
redistribution into specific lines, and averaging over
a 910 -- 1660 \AA\ bandpass, we find an upper limit to the mean
brightness of the continuum fluorescence of 1.1 $\times$ 10$^{-18}$
\bright.  This is a factor of $\sim$ 50 below the background estimate
shown in Figure~\ref{figrock}.  So we find the continuum fluorescence is not
very bright and shows no sign of contributing to the ground
state ro-vibrational population, as there is little in the way of
absorption arising from vibrational levels \vpp\ $>$  2.  This is in
contrast to the strong indication of continuum fluorescence found in the reflection
nebula NGC~2023 by \citet{Meyer:2001}, where they observed absorption lines
from all \vpp\ $\le$ 14 states of \xsig\ in a STIS spectrum of the
central star HD 37903.

The low levels of continuum fluorescence found here are consistent 
with the infrared spectroscopic diagnostics presented by
\citet{Zuckerman:1988}, who found the ratio of S(1) (1--0) / S(1)
(2--1)  (at 2.122 and 2.248 $\mu$m respectively) to be $\approx$ 10,
placing it in the thermal regime well above the value of 2 expected
from fluorescence \citep{Black:1976}.  If the S(1) (1--0)
line were produced by fluorescence, then the 10$^{-4}$
\brightnoaast\ measured by \citet{Zuckerman:1988} would imply a total
ultraviolet brightness $\sim$ 850 times this value (in energy units, 50
times in photon units) and would have easily been detected by the
rocket experiment.
 
The detection of \lya\  fluorescence pumped through two resonant
transitions in the \vpp = 2 vibrational level of the \xsig\ electronic
state (R(6) (1--2) at 1215.730 \AA\ and  P(5)(1--2) at 1216.073) reported
recently by \citet{Lupu:2006},  serves to emphasize that in a radiation
bounded nebula  two-thirds of the \lyc\ photons emitted by the 
star are converted to \lya\ photons \citep{Spitzer:1998}.  The contrast
between the \lya\ and continuum fluorescence is enhanced in our case,
because the number of \lyc\ photons exceeds the number of  photons in
the 912 -- 1120 \AA\ bandpass by a factor of 15.\footnote{Additional
enhancement of \lya\ fluorescence results from the concentration of
excitations into a single \Vp = 1 level of the \bsig\ state, as opposed to the continuum case where
excitations to a multitude of vibrational levels, 0 $\le$ \Vp\ $\la$ 20
for \bsig and 0 $\le$ \Vp\ $\la$ 6 for \cpi, compete for a relatively
smaller number of photons.}  Even so, modeling by \citet{Lupu:2006}
shows the \lya\ fluorescence rate is too low to cause
detectable deviations in a thermalized groundstate population at
a temperature of 2000 K.  We conclude that the observed ro-vibration
population distribution of the hot \Htwo\ is dominated by
collisional processes.  The deviations from a pure thermal distribution
will be addressed in \S~\ref{disrecomb}.

\subsubsection{Molecular Hydrogen Formation and Destruction}

\label{clumpsource}

The observed neutral atomic to molecular hydrogen ratio
($\frac{N(HI)}{N(H_2)}$ = 127) can be used to constrain formation and
destruction processes in the diffuse nebular medium using the formula,
\begin{equation} \label{fd} \frac{dn_2}{dt} = F - D n_2.
\end{equation}   $F$is the sum over all formation processes, $D$ is the
sum over all destruction processes, and $n_2$ is the density of \Htwo.
We will not attempt to provide a comprehensive listing of all the
processes, rather we will make some limiting assumptions to develop the
constraints.

In the limit where formation is absent ($F$ = 0)  the solution for
$n_{2}(t)$  is an exponential with an e-fold time equal to $D^{-1}$.
If we assume the destruction process is radiative dissociation, then
$D=\chi p_{diss}\zeta_{pump}$, where $p_{diss}$ is the mean
dissociation fraction, $\zeta_{pump}$ is the mean probability per
second of an upward radiative transition and $\chi$ the standard
interstellar radiation field multiplier.  \citet{Draine:1996} have
conveniently tabulated $\zeta_{pump}$ and $p_{diss}$ for a number of
transitions for $\chi$ = 1, neglecting self-shielding.  They find
fairly narrow ranges for  $\zeta_{pump}$ and $p_{diss}$ and weighted
mean values of these products over the \Htwo\ ground state
ro-vibrational distribution are relatively insensitive to the actual
distribution for our purposes here. We will use $p_{diss}$ = 0.15 and
$\zeta_{pump}$ = 3.5 $\times$ 10$^{-10}$ s$^{-1}$, yielding $D_1$ = 5.3
$\times$ 10$^{-11}$ s$^{-1}$ as a representative dissociation rate for
$\chi$ = 1.

To compute $\chi$ we use the Rauch
model, which has a total photon luminosity of 1.2 $\times$ 10$^{45}$ ph
s$^{-1}$ in the 912~--~1120~\AA\ bandpass at the stellar surface.  At
0.34 pc, the photon flux  is 8.9 $\times$ 10$^{7}$ ph
cm$^{-2}$ s$^{-1}$, yielding a rather modest scale factor of $\chi$
= 7.4 relative to the average ISM background of 1.2 $\times$ 10$^{7}$
ph cm$^{-2}$ s$^{-1}$ \citep{Draine:1996}.  The resulting e-fold time
for radiative dissociation is  21 years and it would only take 100
years for the $\frac{N(HI)}{N(H_2)}$ ratio to reach its currently
observed level. We conclude the source term for the diffuse medium $F
\neq$ 0, i.e. the \Htwo\ in the diffuse medium must be constantly replenished.

We now turn to the equilibrium limit, i.e.  $\frac{dn_2}{dt} = 0$.  The
formation term is  $F = \gamma n n_1$ where the neutral hydrogen
density is $n_1$.  We consider either formation of \Htwo\ by dust, or by associative detachment $H + H^-   \rightarrow 
H _2  +  e^-$.   In the former case $\gamma_d$ = 3 $\times$ 10$^{-17}$
cm$^3$ s$^{-1}$ \citep{Natta:1998} and $n$ is the gas density.  In the
latter case $n$ is the electron density $n_e$ and $\gamma_-$ is given
by Eq. 15 of \citet{Natta:1998}.  The rate coefficient ($\gamma_-$)
depends explictly  on hydrogen ionization fraction and temperature.  At
$T$ = 2000 K we find $\gamma_-$ $\approx$ $\gamma_d$ when the
ionization fraction $x_+ \equiv$ $n_+/(n_1+n_+)$ = 0.5.  When $x_+$ =
0.05, $\gamma_-$ $\approx$ 15$\gamma_d$.

Equating formation and destruction we solve for the total density $n$
required for formation to keep pace with dissociation. With $\chi$ =
7.4 the dissociation rate is $D$ = 3.9 $\times$ 10$^{-10}$ s$^{-1}$ and
we find, $n=\frac{D}{\gamma}\frac{n_2}{n_1}$ = 10$^{5}$ cm$^{-3}$  for
the dust and the $x_+$ = 0.5 cases.  For  $x_+$ = 0.05 the density is
$\approx$ 7000 cm$^{-3}$.  These densities are consistent with the
\citet{Meaburn:1993} clump estimates for M27, but they are much higher
than the gas and electron densities we found from the \ion{C}{1} analysis in
\S~\ref{cifine} for the transition zone or the electron density given
by \citet{Barker:1984} for the ionized region.

We conclude that the density in the diffuse medium is not high enough
to support formation of \Htwo\ either on dust grains or
through the associative detachment processes and that the most likely source
of excited \Htwo\ is from photo-evaporating clumps
\citep[c.f.][]{Huggins:2002,Lopez:2001}.  Whether the clumps are
reservoirs of \Htwo\ that formed long ago or are currently active in
forming \Htwo\ by these processes is an open question.
The transport process that moves molecular hydrogen out of the
clumpy medium and constantly replenishes the diffuse medium is a dynamical
process, which results in the exposure of \Htwo\ to \lyc\ radiation.

\begin{figure*}
\centerline{\includegraphics[]{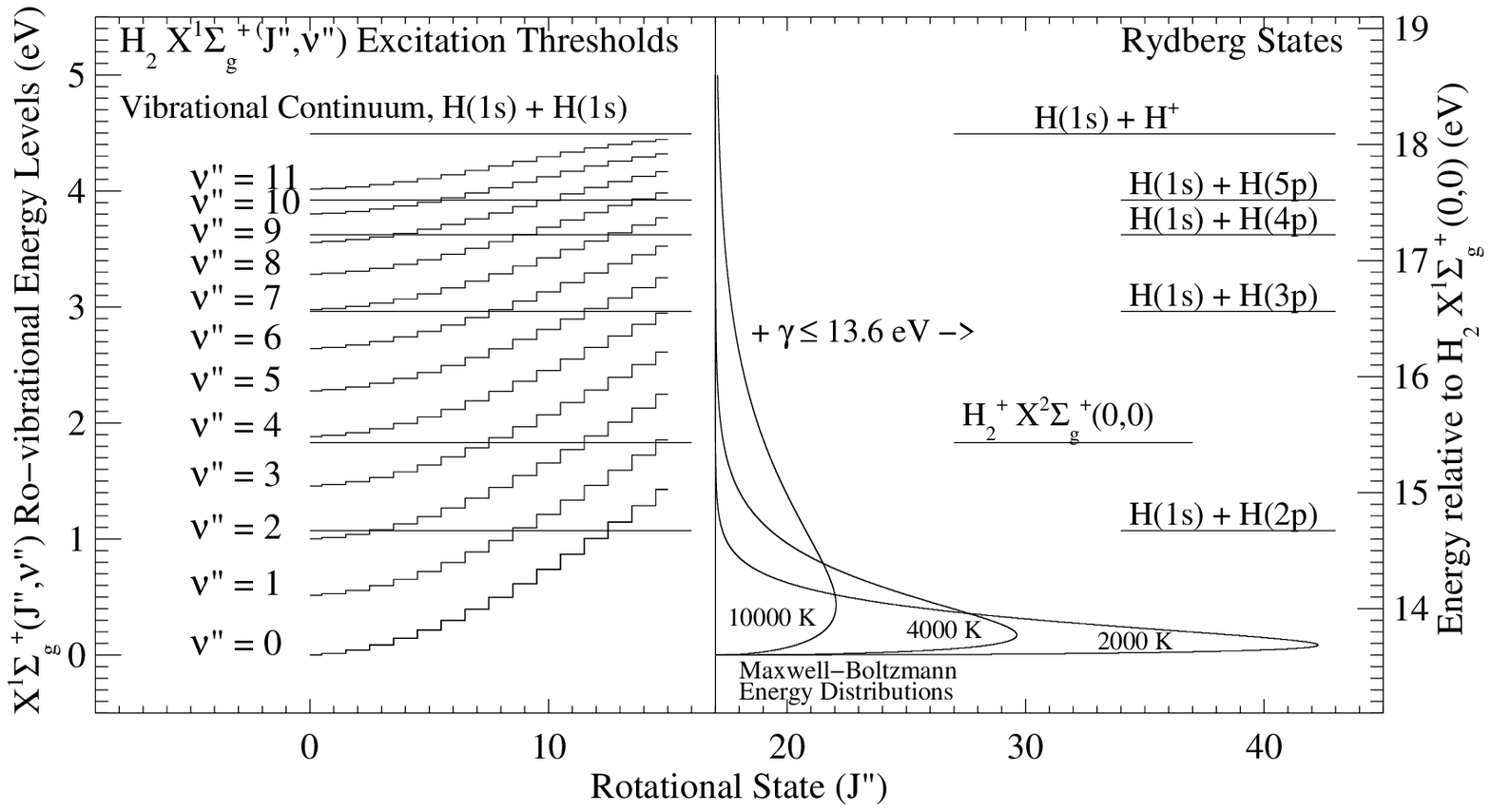}}
\figcaption[/home/stephan/M27/M27redux/elevels.ps]{\label{elevels}
Energy level diagram illustrating the threshold energy ro-vibration levels (\Jpp,\vpp) that can be excited to Rydberg orbitals ($H(1s) + H(nl), 2 \le n \le \infty, l = s, p, d... $) with photons energies $\leq$ 13.6 eV.  The threshold for excitation into the \Htwop\ cation ground state is also shown.  Normalized Maxwell-Boltzmann energy distributions for $T$ = 10000, 4000 and 2000 K show schematically the relative change in the energy level population as a function of temperature. }
\end{figure*}

\subsection{Beyond Spontaneous Dissociation -- \lyc\ Dissociation and Photoionization of \Htwo} 
\label{disrecomb}

We have two outstanding puzzles.  One is the apparent ``two
temperature'' slope in the ro-vibration population
(Figure~\ref{ncolcognew}).  The other is the excess \ha/\hb\ ratio in
the absence of any significant stellar or nebular reddening.  We
speculate here that a confluence of environmental factors -- molecular
hydrogen interacting and excited by a ionized medium that is optically
thin to \lyc\ radiation -- conspire to contribute to the production of
\ha\ emission at a few percent of the radiative recombination rate and
to create the two temperature slope.

We know that while the energies of thermal electrons in the nebula are
high enough to populate the \xsig\ vibrational levels above  \vpp\
$>$ 0, they are not nearly high enough to lead to dissociative
excitation.  For that a radiative process is required such that
$H_{2}(X\,^1\Sigma_{g}^{+}) + \gamma$ ultimately results in  $ H(1s) +
H(3p)$, which can then branch into \lyb\ or \ha\ + \lya. \citet{Stecher:1967} discussed two possible radiative paths
that lead to the dissociation of \Htwo.  In
\S~\ref{h2excite} we described the familiar spontaneous dissociation
process  that leads to $H(1s) + H(1s)$ following excitation into the
Lyman (\bsig) and Werner (\cpi) bands.  The other path (with multiple
channels) is,

\begin{equation}
H_{2}(X\,^1\Sigma_{g}^{+}) + \gamma  \rightarrow
\left\{ 
\begin{array}{ll}   H(1s) + H(nl) &  Pre- and~direct-dissociation, \\ 
H_{2}^{+}(X\,^2\Sigma_{g}^{+}) + e^- & Photoionization,  
\end{array} 
\right .
\end{equation} 
with $n \ge $2 and l = s, p, d ...  the angular momentum of the excited atom. The photoionization channel can further undergo dissociative recombination
with low energy electrons,
\begin{equation}
H_{2}^{+}(X\,^2\Sigma_{g}^{+}) + e^- \rightarrow H(1s) + H(nl).
\end{equation}
These channels are
not usually
considered to be important for \ion{H}{1} regions (PDRs) because the photoionization,
pre-dissociation and direct-dissociation processes all have thresholds
above the Lyman limit.  In such an instance the \Htwo\ is usually cold and optically thick neutral hydrogen shields the \Htwo\
from dissociation \citep{Spitzer:1948}.  However for
\Htwo\ in \ion{H}{2} regions it is not at all clear that
this assumption is justified.  At the interface of a mostly molecular
clump embedded in an ionized medium, as we encounter in M27, there will
be a zone where the \ion{H}{1} and \Htwo\ are optically thin to
ionizing radiation and the \Htwo\ is excited (hot).  In this case the ionization and dissociation of
\Htwo\ by photons with energies above the Lyman limit at 13.598 eV should also be considered.

Direct-dissociation paths proceed through excited ro-vibrational levels
in the \bsig, \bsigp\ and \cpi\ upper electronic states of \Htwo, while
the \bsigpp, \dpi, \dpip\ and \dpipp\ states can under go
pre-dissociation as well as direct-dissociation. The experimental
distinction is that a pre-dissociation process shows structure (in the
absorption cross-sections as a function of wavelength) while
direct-dissociation does not \citep{Lee:1976}.  The structure is a
consequence of the perturbative coupling with the dissociative
continuum of the overlapping excited electronic states.

The vibrational continua of the upper electronic states converge into
Rydberg orbitals,  which are essentially n $\ge$ 2  principal quantum
number electrons in orbital around the cation (\Htwop) core.  At
asymptotic nuclear separation these orbitals become free $H(1s) +
H(nl)$.  Transitions between \xsig\ and its own vibrational continuum
($H(1s) + H(1s)$) at 4.48 eV can proceed only by quadrupole radiation,
while transitions between \xsig\ and the upper singlet electronic states
are dipole allowed.  The vibrational continua of \bsig, \bsigp, and
\cpi\ converge to $H(1s) + H(2l)$ at 14.67 eV (844.8 \AA), \dpi\ and
\bsigpp\ converge to $H(1s) + H(3l)$ at 16.56 eV (748.46 \AA),
\dpip\ converges to $H(1s) + H(4l)$ at 17.22 eV (719.74 \AA) and
\dpipp\ converges to $H(1s) + H(5l)$ at 17.52 eV (707.175 \AA),
measured with respect to (\Jpp, \vpp) = (0,0) of \Htwo\ \xsig.  The
onset of photoionization  (\Htwo\ + $\gamma$ $\rightarrow$ \Htwop\ +
e$^-$) is at 15.43 eV (803.7 \AA) in between the $H(1s) + H(2l)$ and
$H(1s) + H(3l)$.  The right side of Figure~\ref{elevels} shows the energy levels
of Rydberg states in relation to the cation ground state.

\citet{Glass-Maujean:1986} give cross sections for the photo-excitation of
\lya\ ($H(1s) + H(2l)$) from the direct- and pre-dissociation of \Htwo(\xsig).  The total cross
section peaks at the threshold (14.67 eV), reaching 5 $\times$ 10$^{-18}$ cm$^{2}$
and falls monotonically thereafter.  \citet{Cook:1964} have shown the
detailed structure in the \Htwo\ photo-dissociation and
ionization cross section from 550 -- 1000 \AA.  The  photoionization
cross section peaks near 18 eV at $\approx$ 10$^{-17}$ cm$^{2}$.  Some of the discrete dissociation peaks to the red are $\approx$ a
factor of 3 higher at a resolution limit of  0.5 \AA.
\citet{Flannery:1977} have calculated and tabulated cross sections for
excitation out of vibrationally excited states of \Htwo(\xsig,\vpp) +
$\gamma \rightarrow $ \Htwop(\xsigp, $\vp$) + $e^-$ as a function of
wavelength, assuming $\Delta J$ = 0.

\citet{Yan:1998} produced a piecewise continuous function for the
photoionization cross section valid over the interval 300 $\ge E_{ph}
\ge $ 15.4 eV.  At high frequencies the cross section for
\Htwo\ ionization is 2.8 times that for \ion{H}{1}.  Above 18.09 eV
\Htwo\ can undergo dissociative ionization,   \Htwo\ +
$\gamma$ $\rightarrow$ H + H$^+$ + e$^-$, or even  H$^+$ + H$^+$ +
2e$^-$, but the branching ratios into these channels are small over the
wavelengths of \lyc\ emitted by the central star \citep[see][and
references therein]{Yan:1998} where the yield for
dissociative recombination should be $p_{dr}$ $\approx$ 1.

\citet{Stecher:1967} have noted how the thresholds for transition into
the upper level continua are lower for excited ro-vibrational (\Jpp,~\vpp) levels of \xsig.  \citet{Cook:1964} observed this effect in the
redward broadening of the dissociation thresholds by $\approx$ 14 \AA\
in their experiments conducted at room temperature.  It is actually
possible to dissociate and photoionize \Htwo\ with photon energies less
than the Lyman limit, for a high enough ro-vibrational level  (\Jpp,~\vpp).   We quantify this in Figure~\ref{elevels} where the energies of
the ro-vibrational levels in the \xsig\ state in relation to the
asymptotically free states are shown.  Horizontal lines drawn through
the (\vpp, \Jpp) energy levels indicate the thresholds for excitation
with photons $\le$ 13.6 eV in energy.  Maxwell-Boltzmann Energy
distributions with temperatures of 2000, 4000 and 10000 K are drawn to
indicate schematically how the relative populations of the upper levels
change as a function of $T$.

Rotational states above \Jpp\ = 13 in \vpp\ = 0,~\Jpp\ = 9 in \vpp\ =
1, and \Jpp\ = 3 in  \vpp\ = 2  can be excited to $H(1s) + H(2p)$ by
photons with energies $\le$ 13.6 eV.  The steepening of the slope of the level
population in Figure~\ref{ncolcognew} might result from a radiative
depopulation of the higher J'' states into the
asymptotic $H(1s) + H(2p)$ state.  The populations of the upper levels
become selectively depleted, a process we refer to as 
``radiative skimming.''  However, the slope change in the
\vpp\ = 0, 1 curves are observed to occur at at \Jpp\ $\approx$ 6, 8
respectively, which are smaller than given above.  We take this as
evidence for skimming  by \lyc\ photons.  

We also see from Figure~\ref{elevels} that the creation of $H(1s) +
H(3p)$  without the benefit of \lyc\ photons would require
extremely high temperatures  $\gtrsim$ 4000 K.  The two
step process \Htwo(\xsig) + $\gamma \rightarrow $ \Htwop(\xsigp) +
$e^- \rightarrow H(1s) + H(3p)$ is 
energetically more accessible than the direct $H_{2}(X\,^1\Sigma_{g}^{+}) +
\gamma  \rightarrow  H(1s) + H(3p)$.  Observations by \citet{Lee:1976} of
the direct production of Balmer series fluorescence from \Htwo\ by
direct-  and pre-dissociation support this conjecture.  They find the
dissociation yield with respect to total dissociation +
photoionization cross section to be on the order of a few percent.  For
example, at 716 \AA\ the cross section for production of \ha\  was
0.23  $\times$ 10$^{-18}$ cm $^{2}$, which is about a factor of 40 lower than
for the production of \Htwop.

\Htwop\ is also homonuclear \citep{Leach:1995}, so radiative transitions among the 18 vibrational levels of the \xsigp\ ground state are not dipole allowed.
The vibrational continuum of \Htwop\ \xsigp\  ($H^+ + H(1s)$) is only 2.646 eV above the (\Jpp, \vpp) = (0,0) ground state, so electron energy need not be high
to produce dissociative recombination into the overlapping Rydberg states.
Excitation to higher bound states, ($^{2}\Sigma^{+}_{g}$, $^{2}\Pi_{u}$) have poor overlap integrals (Franck-Condon factors) yielding small
branching ratios to these dipole allowed upper levels.
Once created, \Htwop\ will undergo rapid dissociative recombination with free
electrons, \Htwop\ + $e^- \rightarrow H(1s) + H(nl)$, producing a fluorescent cascade of Lyman and Balmer emission.  The amount of \ha\ and \lya\ produced will depend upon the vibrational distribution in the cation energy levels
and the electron impact energies.

\citet{Takagi:2002} has shown in detail how the dissociative
recombination cross sections of the individual $H(nl)$ channels depends
strongly on the vibrational level of the \Htwop\ \xsigp(\vpp) state.
The cross sections are generally highest for e$^-$  collisions with the
lowest energies and fall in proportion to the inverse of the electron energy.
Typical values near zero electron energy range from 10$^{-15}$ --
10$^{-17}$ cm$^{2}$ ignoring rotation.  Including the rotational state
introduces resonances and can strongly modulate the cross
sections at low electron energies.  \citet{Schneider:1994} have
calculated total dissociative recombination rates for \Htwop\ for hydrogen
plasmas with temperatures between 20 -- 4000 K for individual
ro-vibrational states of \Htwop, but they provide no estimate of the
branching ratios into the individual $H(1s) + H(nl)$ final states.

\subsection{Optically Thin Photoionization Destruction of \Htwo\ }

We have suggested that photoionization by \lyc\ photons of excited \Htwo\ in an optically thin medium ultimately results in a
fluorescent cascade of Lyman and Balmer emission.  Verification of
this hypothesis will require exploring the detailed balance of 
hydrogen ions, atoms and molecules in the recombination regime between
Case A (the optically thin limit where Lyman series emissions are not
reabsorbed in the nebula) and  Case B (the optically thick limit where
Lyman series emissions are reabsorbed on the spot).

In the previous section we have given references to the necessary
molecular cross sections for computing the photoionization rates out of
the vibrationally excited molecule and into the excited cation
\citep{Flannery:1977}, along with the cross sections for electron
impact dissociation of excited cation states into the Rydberg manifold
\citep{Takagi:2002}.   The calculation of the  Lyman and Balmer
series emissivity resulting from the two step process of  molecular
photoionization followed by dissociative recombination goes beyond the
scope of this paper.

Here we show an example of the competitiveness of photoionization with
spontaneous dissociation for the destruction of \Htwo\ in the optically
thin limit. The ionization rates at the transition zone (0.34 pc) for
\Htwo\ and \ion{H}{1} were calculated by integrating the cross sections
for these species, using \citet{Yan:1998} for \Htwo\ and \citet[][Eq.
5-4]{Spitzer:1998} for \ion{H}{1}, over the SED of our stellar model
from $\lambda <$ 911.7 \AA\ according to \citet[Eq.
4-36,][]{Spitzer:1998}.  We do not include contributions from direct
and pre-dissociation processes, which are likely to be substantial for
excited \Htwo, so our dissociation estimate will be conservative.  We
find $\zeta_{H2 \rightarrow H2+}$ = 3.4 $\times$ 10$^{-9} $ s$^{-1}$
compared to $\zeta_{H \rightarrow H+}$ = 1.4 $\times$ 10$^{-9}$
s$^{-1}$.  The ratio is 2.4, so we conclude that in the optically thin
limit it is easier to ionize the hydrogen molecule than the atom.

In an optically thin medium at the inferred radius of the transition zone the
molecular photoionization rate $\zeta_{H2\rightarrow H2+}$ exceeds the
far-UV pump rate $\chi\zeta_{pump}$ (\S~\ref{clumpsource}) by a factor of  1.3.  If we assume that the
yield for dissociative recombination (\Htwop\ + e$^-$ $\rightarrow
H(1s)+H(nl)$) following creation of the cation is 100\%, then the
photoionization mediated dissociation of \Htwo\ can be expected
to be  $ p_{dr}\zeta_{H2\rightarrow H2+}$~/~($p_{diss}\chi\zeta_{pump}
) \approx$ 9 times the spontaneous dissociation.  This process
increases the rate required for the clumps to resupply the diffuse
medium with \Htwo\ in the steady state limit.  We note that
\citet{Natta:1998} do not include \lyc\ dissociation of \Htwo\ in their model.  This may be why they find the shell mass is dominated by \Htwo\ as opposed our finding that $\frac{N(HI)}{N(H_2)} \gg $ 1.

\section{Conclusions and Suggestions for Future Investigations}

M27 is an excellent laboratory for testing theories of formation,
excitation and destruction of molecules and dust, because the stellar temperature, mass, gravity, as well as the nebular
distance, masses, abundances and excitation states, are well quantified.

We find that the diffuse nebular medium is generally inhospitable to
molecules and dust.  The diffuse \Htwo\
accounts for a small fraction ($\sim$ 0.3 \%) of the total nebular
mass.  \Htwo\ is easily destroyed by the stellar radiation
field in the diffuse medium on a timescale $\sim$ 20 years.  A steady
state abundance requires a source, most likely the clumpy medium.   If
the clump density is high enough, they may support molecular
hydrogen formation via dust or mediated by radiative detachment, although we cannot rule out photoevaporation of relic material.

The finding of neutrals and molecules at a transition velocity between
slow-high and fast-low ionization material suggests collisional
interaction with charged particles is the most important process for
establishing the observed ro-vibration level populations in molecular
hydrogen.  \citet{Lupu:2006} finds the rate of \lya\ pumping is not
high enough to change the level population of a thermal distribution
and we find the rate of continuum pumped fluorescence is even lower.
However, if the neutral hydrogen is optically thin in
the transition zone, then stellar \lyc\ can radiatively skim the upper
rotational states of \vpp\ = 0, 1 and create the apparent break in the high and low temperature slopes at the observed levels, but this is
likely to be a second order process.

Dissociation and ionization of the hydrogen molecule by \lyc\ photons
can also lead to the production of Lyman and Balmer series emission and
may provide an explanation for the excess \ha/\hb\ ratio observed in
the apparent absence of extinction by dust.  In the optically thin
limit \lyc\ mediated destruction rates can exceed by an order of
magnitude the spontaneous dissociation rate and should be considered 
when calculating the equilibrium abundance of excited \Htwo\ in \ion{H}{2} regions.

At this late stage in the evolution of M27 most of the original AGB
atmosphere has been dissociated and accelerated to velocities in excess
of 33 \kms\ but not above 68 \kms.  A high speed radiation driven
stellar wind is not detected in the present epoch. The mechanism for
acceleration remains unidentified.

The analysis of the formation and destruction of \Htwo\
rests in part on the determination of the \ion{H}{1} velocity
structures identified in the Dwingeloo survey data to guide the column
density determination carried out with the STIS and \FUSE\ data.
However, our conclusions regarding the overall abundance of molecular
hydrogen with respect to \ion{H}{1} and the need for a molecular source
for the diffuse medium are immune to changes in the atomic abundance by
factors of 10.  Nevertheless it would be prudent to check the velocity
structure found here with a high resolution mapping using a telescope
with a smaller beam ($\approx$ 4\arcmin).  It appears that the velocity
separation of M27 permits discrimination with respect to the galactic
\ion{H}{1} background.  This and a reasonable signal strength have
allowed the apparent detection of \ion{H}{1} in a PN where only upper
limits have existed previously \citep{Schneider:1987}.

Our lowest bound for $N(CO)/N(H_2)$ $<$ 10$^{-3}$ in the diffuse nebular medium, is a factor of 30 higher than ratio used by \citet{Huggins:1996} to constrain the CO clump mass.  Should a \HST\ servicing mission 4 be successful in resurrecting STIS then higher signal-to-noise data could be acquired to examine the  $N(CO)/N(H_2)$ more carefully, using a number of rovibrational bands.  It should be possible to place limits on $\log{N(CO)} \lesssim $ 13   \citep[e.g.][]{Burgh:2007}.

We emphasize that our analysis applies to the direct line of sight gas,
the diffuse nebular medium,  which we find to be dust free.  The
existence of clumps with column densities of CO in excess of 10$^{16}$
cm$^{-2}$ immediately adjacent to regions with CO column densities
$\lesssim$ 10$^{14}$ cm$^{-2}$ attests to the steep density gradients
within the nebular environment and the probable low covering fraction
of the densest clumpy medium.  The clumps may well be reservoirs of
material depleted from the diffuse medium.  Measuring the abundance of
metals on either side of the neutral transition velocity could provide
information on whether photo-evaporation of the globules is an
important process for the enrichment of metals in the high velocity
zone \citep{McCandliss:2007}.  In addition, infrared spectral and imaging
observations with Spitzer could tell us much more about the prevalence
of dust and molecules elsewhere in the nebula.  Mapping of the nebular
Balmer emission line ratios with respect to the \Htwo\ S(1)
(2--1) / S(1) (1--0) emission ratio would be especially useful for further
exploring the connections between \Htwo\ production of  Balmer emission
and the prevelence of thermal or fluorescence processes in the clumpy
medium.

During the review of this paper, work by \citet{O'Dell:2007} was brought to our attention.  O'Dell et al. derive the total
flux in the infrared \Htwo\ emission lines in the Helix Nebula and
argue that only \lyc\ radiation from the star has enough power to
sustain the \Htwo\ IR emission at the observed levels.  They also
propose that charge exchange between \Htwo\ and O$^+$ could be a
significant dissociation channel.  Future work examining the efficiency
of their proposed charge exchange mechanism for mediating the
dissociation of \Htwo\ in comparison to the Lyc
dissociation  process proposed here, would be a useful addition to the
nascent field of Lyc dominated PDR's.

\acknowledgments

We are grateful to Patrick J. Huggins who suggested that the molecular
hydrogen kinematics might share similarities with CO and providing
encouragement to complete this work. We acknowledge the sounding rocket
mission operation teams from Wallops Flight Facility and the Physical
Science Lab operated by New Mexico State University and located in
WSMR.  The coordination of observing specialists Jack Dembicky, Russet
McMillan and Gabrelle Saurage at APO allowed for the efficient
execution of an ambitious observing program.  Based on observations
made with the NASA-CNES-CSA Far Ultraviolet Spectroscopic Explorer.
\FUSE\ is operated for NASA by the Johns Hopkins University under NASA
contract NAS5-32985.  Observations were also obtained with the Apache
Point Observatory 3.5-meter telescope, which is owned and operated by
the Astrophysical Research Consortium. In addition some of the data
presented in this paper were obtained from the Multimission Archive at
the Space Telescope Science Institute (MAST). STScI is operated by the
Association of Universities for Research in Astronomy, Inc., under NASA
contract NAS5-26555.  Support for MAST for non-HST data is provided by
the NASA Office of Space Science via grant NAG5-7584 and by other
grants and contracts.



Facilities: \facility{\FUSE}, \facility{\HST} (STIS),  \facility{\IUE}, \facility{WFF}, \facility{WSMR}, \facility{Dwingeloo}, \facility{APO} (DIS).








\end{document}